\documentclass[conference]{IEEEtran}

\usepackage{amsmath,graphicx,epsfig,float,subfigure,times,latexsym,verbatim,mathrsfs,amssymb,textcomp,txfonts,color,mathrsfs}
\usepackage[flushleft]{threeparttable}
\usepackage{algorithm}
\usepackage{algorithmic}
\usepackage{url}

\IEEEoverridecommandlockouts

\ifCLASSINFOpdf
\else
 \fi

\begin{document}
%

\title{\huge User Capacity of Wireless Physical-layer Identification:\\An Information-theoretic Perspective \vspace{-10pt}}

\author{\IEEEauthorblockN{
\thanks{Zhi Sun's research is supported in part by the US NSF under Grant CNS-1547908. Kui Ren's research is supported in part by the US NSF under grant CNS-1318948. Wenhao Wang's jointly training PhD progroam is supported by China Scholarship Council.}
\textbf{Wenhao Wang}\IEEEauthorrefmark{1}\IEEEauthorrefmark{2}, \textbf{Zhi Sun}\IEEEauthorrefmark{2}, \textbf{Kui Ren}\IEEEauthorrefmark{3}, \textbf{Bocheng Zhu}\IEEEauthorrefmark{1} 
\IEEEauthorblockA{\IEEEauthorrefmark{1}
School of Electronics Engineering and Computer Science, Peking University, \\Beijing 100871, China, Email: \{wenhaowang, zhubc\}@pku.edu.cn}
\IEEEauthorblockA{\IEEEauthorrefmark{2}
Department of Electrical Engineering, \IEEEauthorrefmark{3}Department of  Computer Science and Engineering, \\University at Buffalo, The State University of New York, \\Buffalo, New York 14260, USA, E-mail: \{wenhaowa, zhisun, kuiren\}@buffalo.edu}
}\vspace{-15pt}}

\maketitle

\begin{abstract}

Wireless Physical Layer Identification (WPLI) system aims at identifying or classifying authorized devices based on the unique Radio Frequency Fingerprints (RFFs) extracted from their radio frequency signals at the physical layer. Current works of WPLI focus on demonstrating system feasibility based on experimental error performance of WPLI with a fixed number of users. While an important question remains to be answered:  what's the user number that WPLI can accommodate using different RFFs and receiving equipment. The user capacity of the WPLI can be a major concern for practical system designers and can also be a key metric to evaluate the classification performance of WPLI. In this work, we establish a theoretical understanding on user capacity of WPLI in an information-theoretic perspective. We apply information-theoretic modeling on RFF features of WPLI. An information-theoretic approach is consequently proposed based on mutual information between RFF and user identity to characterize the user capacity of WPLI. Based on this theoretical tool, the achievable user capacity of WPLI is characterized under practical constrains of off-the-shelf receiving devices. Field experiments on classification error performance are conducted for the validation of the information-theoretic user capacity characterization.

\end{abstract}
\vspace{-5pt}

\IEEEpeerreviewmaketitle

\section{Introduction}
\begin{figure}
\centering
\includegraphics [width=2.0in] {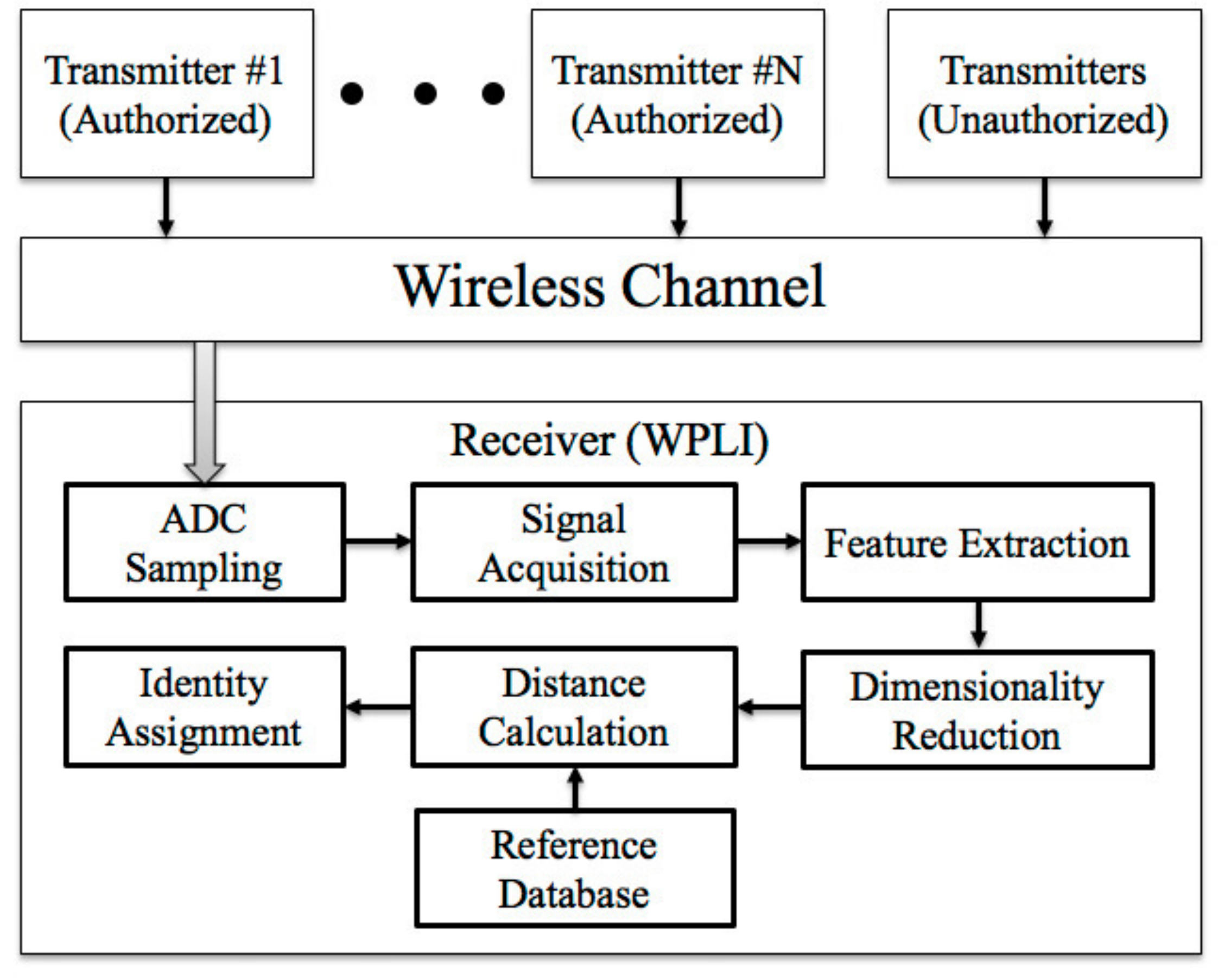}
\vspace{-5pt}
\caption{Typical logic procedures and application scenarios in WPLI.}
\vspace{-15pt}
\label{fig:system_overview}
\end{figure}

Wireless Physical-layer Identification (WPLI) is a promising wireless security solution. Since the software-level device identities (e.g., IP or MAC address) can be manipulated, the physical layer feature cannot be modified without significant efforts. The physical layer features are extracted from signal by WPLI to form the radio frequency fingerprints (RFFs) which are rooted in the hardware imperfections of analog (radio) circuitry at the transmitter device \cite{danev2012physical}. Fig.\ref{fig:system_overview} illustrates the processing procedures of WPLI and typical application scenarios. The signals are obtained by the identification system through an acquisition setup to acquire signals from devices. A feature extraction module is then to obtain selected kinds of identification-relevant feature from the identification signal to form a fingerprint. A fingerprint matcher compares the fingerprints with
reference fingerprints stored in database using dimensionality reduction classification technique. The identities are classified and assigned to devices. Two application scenarios are involved in WPLI: (i)  identification scenario is between all unauthorized imposters and the whole authorized users. (ii) classification scenario is the N-class identification between all authorized users within this network \cite{danev2012physical, danev2009physical}.

     The RFFs are rooted in the hardware imperfections of the transmitter device \cite{danev2012physical}, which include the nonlinearity of RF front-end system \cite{polak2011identifying}, \cite{polak2011rf}, \cite{liu2008specific}, clock jitter \cite{zanetti2010physical, jana2010fast},  distortions due to modulator sub-circuit \cite{brik2008wireless}, etc. The classification procedure of WPLI is to use RFF features to estimate the user identities. However, due to channel effects, in-band hardware noise, and resolution errors of extraction algorithms, the features extracted by receiver become random variables with certain distribution which brings the uncertainty to the classification results. 
 If more user classes are kept adding into WPLI, the distributions of features of different identities are more likely to overlap. Hence the uncertainty between feature and identity is increased, resulting the increase of classification errors. Current works in WPLI research area, mainly focus on demonstrating feasibility of system with the classification error performance of a fixed-number network measured by high quality receiving equipment. For instance, in \cite{danev2009transient}, 50 COTS Tmote Sky nodes and an oscilloscope are utilized to achieve a high sampling rate (4GS/s). In \cite{brik2008wireless}, 138 Network Interface Cards (NIC) are measured and a vector signal analyzer are used as the receiver. In \cite{scanlon2010feature}, 54 Universal Mobile Telecommunications System (UMTS) user equipment (UE) devices and a signal spectrum analyzer are utilized.  To our best knowledge, no existing works have analyzed the user number that WPLI can accommodate using different RFFs and receiving equipment within certain performance, i.e., the user capacity of the WPLI.

Since the existing research analyses are mainly conducted in different experiment scenarios, the results of which cannot have the repeatable accuracy due to different individual experiment setup. To this end, in this paper, information-theoretic analyses are utilized to provide the theoretical tool, which can be universally applied for various types of WPLI. This theoretical tool can be a fundamental approach to describe the uncertainty between feature and user class identity in WPLI. Specifically, entropy can be used as a measure of the uncertainty on the values taken by a feature member. Meanwhile mutual information can be seen as the reduction in the uncertainty of one feature member due to the knowledge of user class identity. \cite{scanlon2010feature,brown2009information}. Moreover, the classification error performance is restricted by the uncertainty remains in the feature member after the reduction of mutual information \cite{cover2012elements}. Based on the relations of uncertainty, the key factors in WPLI, including feature, class identity, classification error performance, and user capacity, can be jointly analyzed. However, to date, no existing work in the research field of WPLI has fully covered this research direction.

In this paper, we establish a theoretical understanding of user capacity of WPLI in an information-theoretic perspective. Based on mutual information of RFF, an information-theoretic approach is established to characterize the user capacity of WPLI. The RFF feature of WPLI is modeled according to the signal processing procedures of WPLI. The mutual information between RFF feature member and user identity is calculated. The ensemble mutual information (EMI) between RFF and identity is obtained using an approximation calculation. The user capacity of WPLI is derived using the EMI and class identity entropy. To illustrate the usage of this theoretical tool, we use a experiment-based approach to calculate the mutual information and then to derive the achievable user capacity under practical constrains of different application cases. Experiments on classification error performance of a practical system are also conducted to validate the user capacity characterization for each application case setting.

\section{Information-theoretic Analyses of User Capacity}
\label{Modeling}
In this section, we firstly provide the information-theoretic modeling of RFF feature according to the processing procedures. The mutual information between fingerprinting feature member and user identity is modeled. The model to calculate ensemble mutual information between RFF and identity is then given. Finally, the user capacity is derived using ensemble mutual information and entropy.

\subsection{Modeling of RFF feature}
The beginning of the RRF classification procedure is the  Analog-to-Digital Converter (ADC) sampling procedure of received signal. This signal can be either baseband signal or passband signal resulting in different signal type are utilized to extract the fingerprints which are basedband preamble \cite{zanetti2010physical,suski2008using} or passband transient signal respectively \cite{danev2009transient, barbeau2006detection}. The ADC sampling procedure can be modeled as,
\begin{gather}
\label{eq: ADC}
\textstyle s[n]=s(nT_s)+\eta+\xi_{ADC},
\end{gather}
where $s(t)$ is the received analog signal, $s[n], n\in\mathbf{N}$  is the sampled digital signal, $\mathbf{N}$ is the set of all sampled digital signals in this round of identification. $\eta$ is the in-band AWGN noise which can be measured by the receiver. $\xi_{ADC}$ is the random ADC quantization error,  for Q-bit ADC quantization and input dynamic range $U$ Vp-p, the maximum quantization error is $ \delta_{ADC}=2^{-Q}U$.

After ADC sampling procedure, the signal sequence then go through the signal acquisition procedure to extract the valid part of the signal i.e., the preamble or transient part. The next procedure is feature extraction to obtain the fingerprinting feature from the signal, which can be modeled as,
\begin{gather}
\label{eq: feature}
\textstyle \mathbf{X}_{1:M}=Feature(\mathbf{S}_{1:N}),
\end{gather}
where $\mathbf{S}$ is $N$ point raw signal data vector set, $\mathbf{X}$ is the fingerprint feature set with the feature dimensionality $M$. The feature dimensionality depends on the feature selection approach. For instance, the spectral feature in \cite{danev2009transient} is high-dimensional feature of which the $M$ is the number of FFT points. In \cite{zanetti2010physical}, two single-dimensional features, TIE error and average signal power, are utilized as a combined feature. While in \cite{brik2008wireless}, a combined low-dimensional feature of frequency error, I/Q offset, magnitude error and phase error are used as fingerprinting feature. Since the multi-dimensional feature are widely applied in the feature selection, dimensionality reduction techniques are applied to reduce the computation burden and find more discriminant subspaces which highlight the relevant features that may be hidden in noise \cite{danev2012physical}. Currently typical dimensionality reduction techniques in machine learning are applied for WPLI classification scenario, including PCA \cite{danev2009physical}, Fisher LDA \cite{danev2009transient}, Maximum Mutual Information (MMI) \cite{scanlon2010feature} etc.

 \subsection{Mutual information between RFF and identity}
To derive the user capacity $N_C$, we firstly calculate the mutual information between RFF feature and its identity. Specifically, variable $X$ is the one single dimensional component of the feature vector i.e., $X\in \mathbf{X}$. Y denotes the user class identity of this RFF feature. The value of $X$ and $Y$ varies for each testing RFF sample received by WPLI. Consequently, the number of $X$ values, $N_X$, equals the number of all the test samples received by WPLI. The entropy of feature values can be calculated as, $ H(X) =-\sum_{i=1}^{N_X}p(x_i)\log(p(x_i)) $.  While the number of $Y$, $N_Y$ is the number of users connected to WPLI. The classification procedure of WPLI is to utilize larges samples of $X$ with different identities to decide the user identity $Y$.  Hence the conditional entropy, which describe the uncertainty remaining in $X$ after obtained the outcome of $Y$, can be calculated as $H(X|Y) =-\sum_{j=1}^{N_Y} p(y_j) \sum_{i=1}^{N_X}p(x_i|y_i)\log(p(x_i)|y_i)$. The mutual information between $X$ and $Y$ can be finally derived as,
 \begin{align}
\label{eq: MI}
\textstyle I(X;Y)&=I(Y;X) =H(X)-H(X|Y)
\\\textstyle &=\sum_{j=1}^{N_Y} \sum_{i=1}^{N_X} p(x_i y_i)\log(\frac{p(x_i y_i)}{p(x_i)p(y_j)}). \notag
\end{align}

\begin{figure}[!t]
  \centering
  \subfigure[]{
    \includegraphics [width=1.6in] {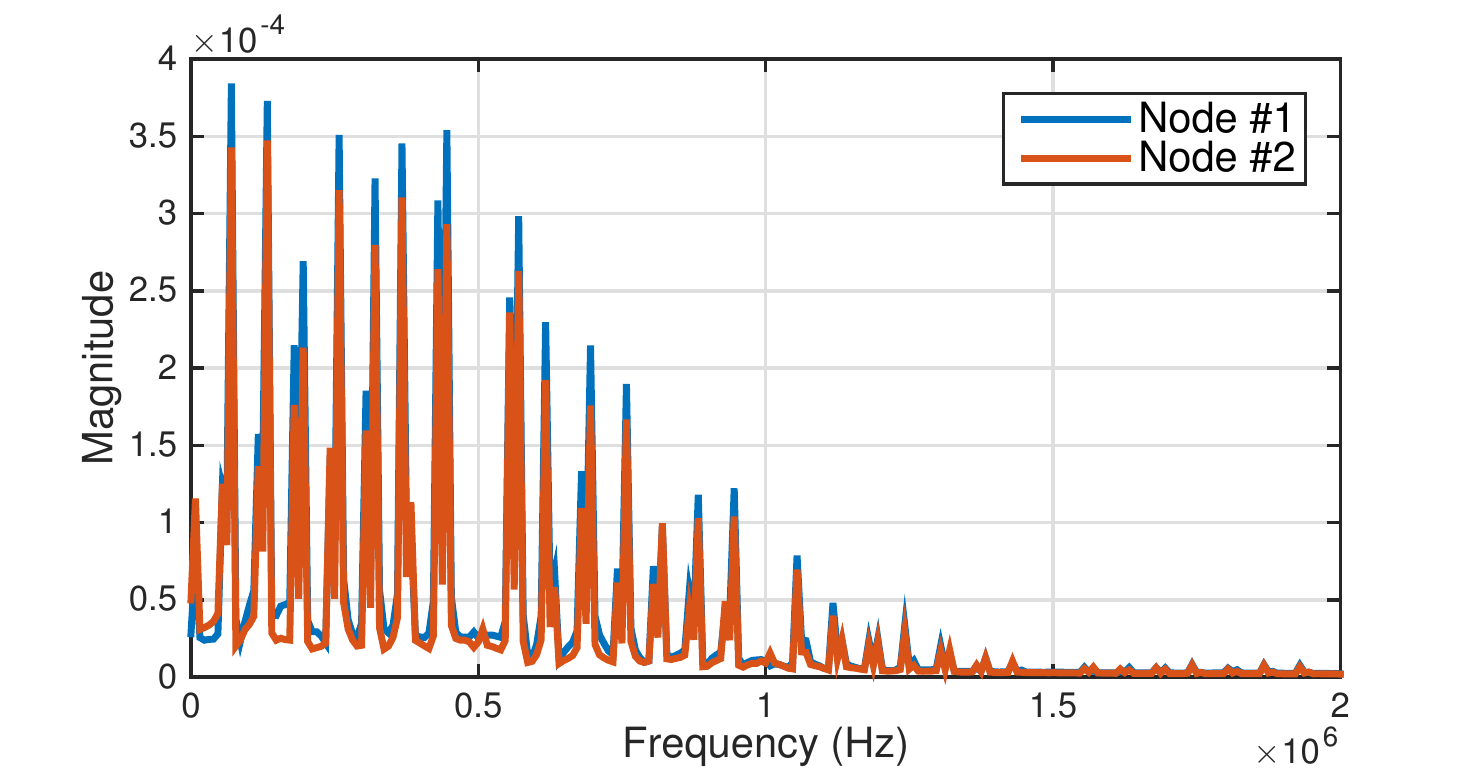}
    \label{fig_PSD}}
  \subfigure[]{
    \includegraphics [width=1.6in] {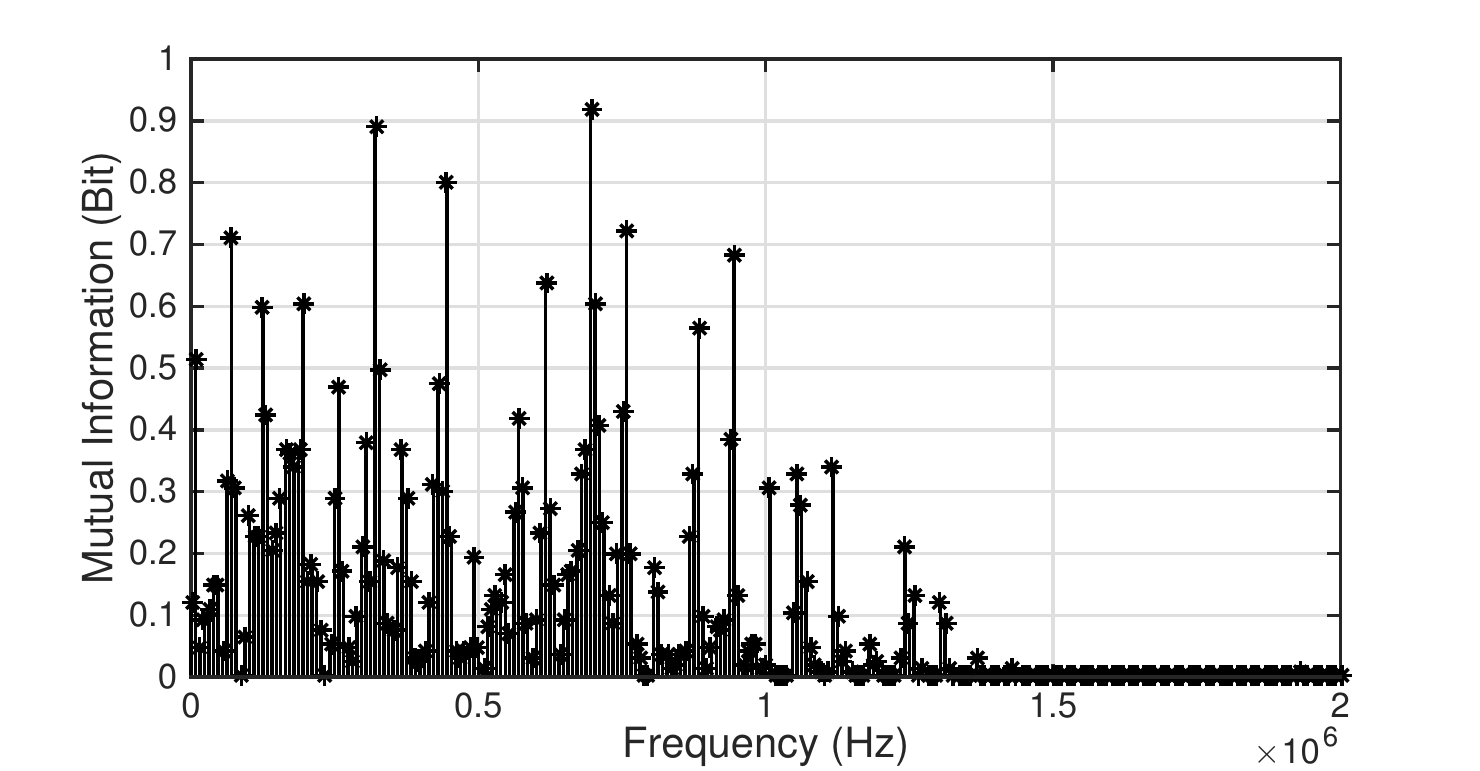}
    \label{fig_MI}}
  \caption{ (a) Preamble PSDs of Micaz sensor nodes.  (b) Mutual information between spectral feature member and identity.}\vspace{-15pt}
     \label{fig_PSD_MI}
\end{figure}

 For instance, if the whole signal spectrum is used as feature $\mathbf{X}$ for RFF, each value of frequency can be seen as the one variable $X$ for the spectral feature. $x_i$ is the specific magnitude value of each frequency point is of the $i$th RFF sample. Hence mutual information between each frequency point and identity can be measured and calculated using large number of tests. The specific measure and calculation approach which will be detailed discussed using an application case in section \ref{sec: app_case}. To have a clear understanding of mutual information between feature member and identity, in Fig.\ref{fig_PSD_MI} we present the two PSDs of signal preambles of two Micaz sensor nodes and corresponding mutual information between each frequency point and signal identity. The difference of spectrums of different devices is the reason that spectral feature can be utilized to classify the identities. From the figures, we can see that the frequency points which have larger differences in spectrum also show larger mutual information value with their identity. Just as the discussion in \cite{scanlon2010feature} and \cite{brown2009information}, mutual information is a significant metric to characterize the relevance of the feature to its identity.

If only single dimensional feature is utilized to form a RFF, the mutual information between RFF and identity can already be calculated using equation (\ref{eq: MI}).  However, as our previous discussion, even single dimensional features are combined to form a multi-dimensional feature $\mathbf{X}$ to form a RFF. Hence the Ensemble Mutual Information (EMI) between ensemble feature and class identity, $I(\mathbf{X};Y)$, needs to be calculated to characterize the relation between the ensemble RFF and the identity. In \cite{brown2009information, zhou2010multi}, a definition for the ensemble mutual information is given. However, as the increasing of dimensionality of ensemble feature, exponential number of possible variable values will make the calculation for this approach infeasible in practice. In \cite{ozertem2005detection,ozertem2006spectral}, the EMI between high-dimensional feature and class identities is given as,
\begin{align}
\label{eq: EMI}
\textstyle  I(\mathbf{X};\it Y) &=\sum_Y \int p(\mathbf{x}, y) \log \frac{p(\mathbf{x}, y)}{p(\mathbf{x}) p(y)} d\mathbf{x}
\\ \textstyle &=\sum_Y p(y) \mathbb{E}_{\mathbf{x}| y} \left [ \log \frac{p(\mathbf{x}| y)}{p(\mathbf{x})} \right ], \notag
\end{align}
The pdfs $p(\mathbf{x}| y)$, $p(\mathbf{x})$ and the conditional expectation $\mathbb{E}_{\mathbf{x}| y} $ can be  approaximatedly calculated using nonparametrical Kernel Density Estimator (KDE) with $K(.)$ as the kernel \cite{fasshauerkernel}, 
\begin{align}
  \label{eq: EMI_app}
 \textstyle I(\bf{X};\it Y) & \approx\sum_Y \frac{p(y)}{N_Y} \sum_{j=1}^{N_Y} \log \frac{(1/N_Y)\sum_{i=1}^{N_Y} K(\mathbf{x}_j^y-\mathbf{x}_i^y)}{(1/N_\mathbf{x})\sum_{i=1}^{N_Y} K(\mathbf{x}_j^y-\mathbf{x}_i)}
 \\ \textstyle &\approx \sum_Y \frac{p(y)}{N_Y} \sum_{j=1}^{N_Y} \log \left [ \frac{\bar{\varphi}^T(\mathbf{x}_j)\bar{\Lambda} \mathbf{\bar{\mu}}_y }{\bar{\varphi}^T(\mathbf{x}_j)\bar{\Lambda} \mathbf{\bar{\mu}}} \right ] \notag
\end{align}
where the kernel $K(.)$ can be calculated with the eigenvectors  $\mathbf{\varphi(x)}$ and the eigenmatrix $\mathbf{\bar{\Phi}_x}=[\mathbf{\varphi(x)}_1, ...,\mathbf{\varphi(x)}_N  ]$,  $ \mathbf{\bar{\mu}}_y=(1/N_Y)\mathbf{\bar{\Phi}_x}\mathbf{m}_y$ is the average eigenvector for class $y$, $ \mathbf{\bar{\mu}}=(1/N_\mathbf{x})\mathbf{\bar{\Phi}_x}\mathbf{1}$ is the average eigenvector for all the training samples, and $N_\mathbf{x}$ is the number of all RFF feature samples. 

\subsection{User capacity of WPLI}
    The WPLI system finally assign the identity of testing RFF to the class with minimal feature distance scores between reference RFFs. After a large number of sample tests, the WPLI classification performance can be evaluated using average classification error rate as the metric \cite{danev2009physical, zanetti2010physical} which can be denoted as $P_e$. With the EMI $I(\mathbf{X};Y)$ obtained, the important property of mutual information related to classification error rate $P_e$ can be utilized to derive the user capacity of WPLI. In \cite{brown2009information,cover2012elements,zhou2010multi}, the information-theoretic bounds for classification error rate are given in details using Fano's Inequality (note that the inequality is valid for three or more classes scenario). Hence the classification error rate of WPLI can be bounded as,
 \begin{align}
  \label{eq: bounds}
\textstyle \frac{H(Y)-I(\mathbf{X};Y)-H(P_e)}{\log(N_Y-1)}
\leq P_e 
\leq \frac{1}{2} \left (H(Y)-I(\mathbf{X};Y) \right).
\end{align}
These bounds determine that no classifier can possibly achieve better than error lower bound and also there exists a classifier that can achieve at least error upper bound. The bounds are restricted by two terms. One is the ensemble mutual information between feature and identity $I(\mathbf{X};Y)$, the other is class identity entropy $H(Y)$. Considering a specific scenario of WPLI, the stable mutual information between feature and identity, can be measured with a large number of test samples \cite{scanlon2010feature} and the EMI can be calculated using equation (\ref{eq: EMI}). Hence the bound error rate can be bounded by the class identity entropy which is directly related to the user number $N_Y$ of WPLI. Considering an equal-identity-probability WPLI system, i.e., $H(Y)=\log(N_Y)$, the upper-bound user capacity can then be derived as,
 \begin{gather}
  \label{eq: up_UC}
\textstyle N_C= \max(\mathbf{N}_Y) | \frac{\log(N_Y)-I(\mathbf{X};Y)-H(\lambda)}{\log(N_Y-1)}\leq\lambda,
\end{gather}
where $N_C$ is the user capacity, $\mathbf{N}_Y$ is the set of all possible user number, $Y$ is the user identity, and $\lambda$ is the performance threshold for classification error rate $P_e$. By far, the theoretical tool to derive the user capacity of WPLI is given.

\section{Application Case Study on User Capacity under Practical Constrains}
\label{sec: app_case}
To apply this theoretical tool, we conduct an application case study to illustrate its usage for a specific type of WPLI. We use a experiment-based approach to calculate the mutual information between feature and identity. Then the achievable user capacity of this type of WPLI is derived under practical constrains of different application case settings. Moreover, the effects of key system parameters on user capacity are evaluated and analyzed.

\subsection{Case overview}

The most existing works try to present the best performance with the high quality receiving equipment. Differently, we try to derive the user capacity and evaluate the system feasibility using the existing approach under practical constrains of off-the-shelf devices. The details of this application case are given as,
\begin{enumerate}[\IEEEsetlabelwidth{8)}]
\item Feature selection: FFT spectrum of baseband preamble \cite{danev2009physical,zanetti2010physical,scanlon2010feature,suski2008using}.
\item Transmitter: Micaz, Imote2 and TelosB sensor nodes (3 typical models with the same ZigBee Protocol radio chip). 
\item Receiver: USRP N210 with SBX daughter board (14-bit ADC).
\item Sampling rates: 2M$\sim$10MS/s
\item Number of FFT points: 64$\sim$2048p.
\item Communication channel: indoor AWGN channel (SNR=20$\sim$30dB).
\item Number of transmitters (user class identities): 40 in all.
\item Number of signal samples: 2000 samples per class.
\end{enumerate}

\subsection{User capacity characterization}

After the one-time collection of the raw signals from sensor nodes, the training samples of RFF can be obtained. We calculate the ensemble mutual information for each RFF and its class identiy using the nonparametrical Kernel Density Estimation approach in equation (\ref{eq: EMI_app}). With the EMI obtained, the next step is to characterize the user capacity using equation (\ref{eq: up_UC}). Subsequently, we try to derive the user capacity under different constrains of key parameters in the user capacity modeling and RFF forming, including number of training transmitters $N_Y$, in-band AWGN noise level, ADC quantization bits $Q$, number of FFT points $N_{FFT}$ and sampling rate of receiver $f_s$. we set the parameters due to different typical application scenarios of WPLI and derive the user capacity with targeted performance as the following figures. In each case, we present the ensemble mutual information (EMI) (blue curve) together with user capacity under 1\% and 10\% classification error rate i.e., $N_C|P_e\leq1\%$ (black curve) and $N_C|P_e\leq10\%$ (red curve). 

\subsubsection{Effects of number of training transmitters}

\begin{figure}[!t]
  \centering
    \includegraphics [width=2.0in] {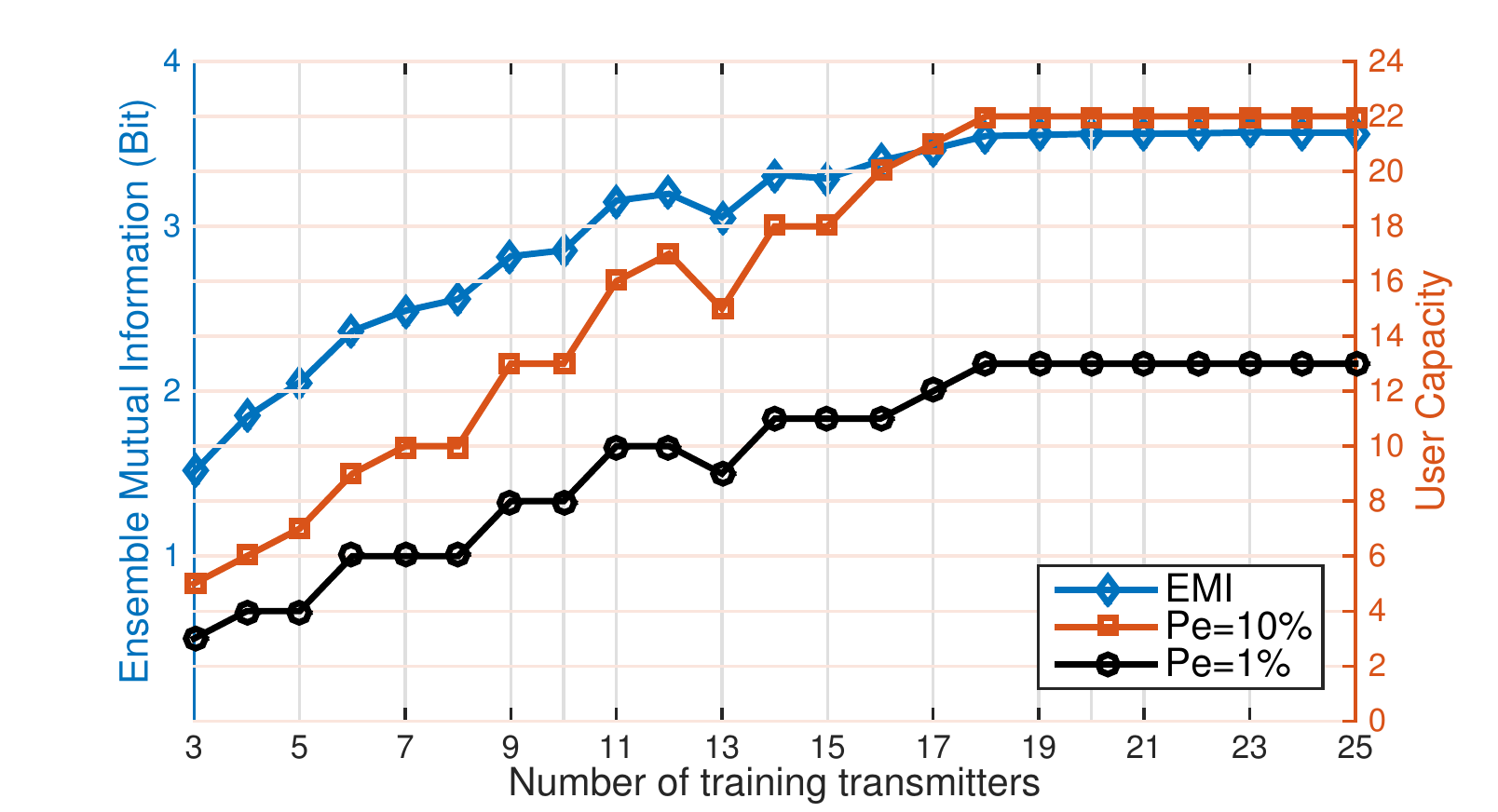}
  \caption{ User capacity under different numbers of training transmitters.}\vspace{-15pt}
     \label{fig_num_nodes}
\end{figure}

Since we obtain the raw RFF samples from limited number of transmitters and limited number of RFF samples to characterize the user capacity, firstly, we characterize user capacity results using RFF samples from 3 to 40 training transmitters to find out what's the least number of class identities we need to characterize a stable user capacity for this system. We collect the raw RFF samples in a typical system setting as $f_s=4MS/s$, $N_{FFT}=512p$, and $SNR=24dB$. In Fig. \ref{fig_num_nodes}, the user capacity is presented from 3 to 25 training transmitters together with the EMI we obtained from these samples. The EMI directly related to classification error rate and user capacity, the relation of which can be easily observed in the figures. In the beginning, when the number of training transmitters is too small, the obtained EMI is also small which results in the user capacity is near to $N_Y$. As the increasing of $N_Y$, the results are increased unstably. After the number of training transmitters is larger than 18, the characterization result becomes a stable and reliable result which is $13|P_e\leq1\%$ and $22|P_e\leq10\%$. Hence in order to characterize the user capacity we at least should use 13 nodes to collect the raw training RFF samples.

\subsubsection{Effects of RFF noise level}

\begin{figure}[!t]
  \centering
  \subfigure[]{
    \includegraphics [width=1.6in] {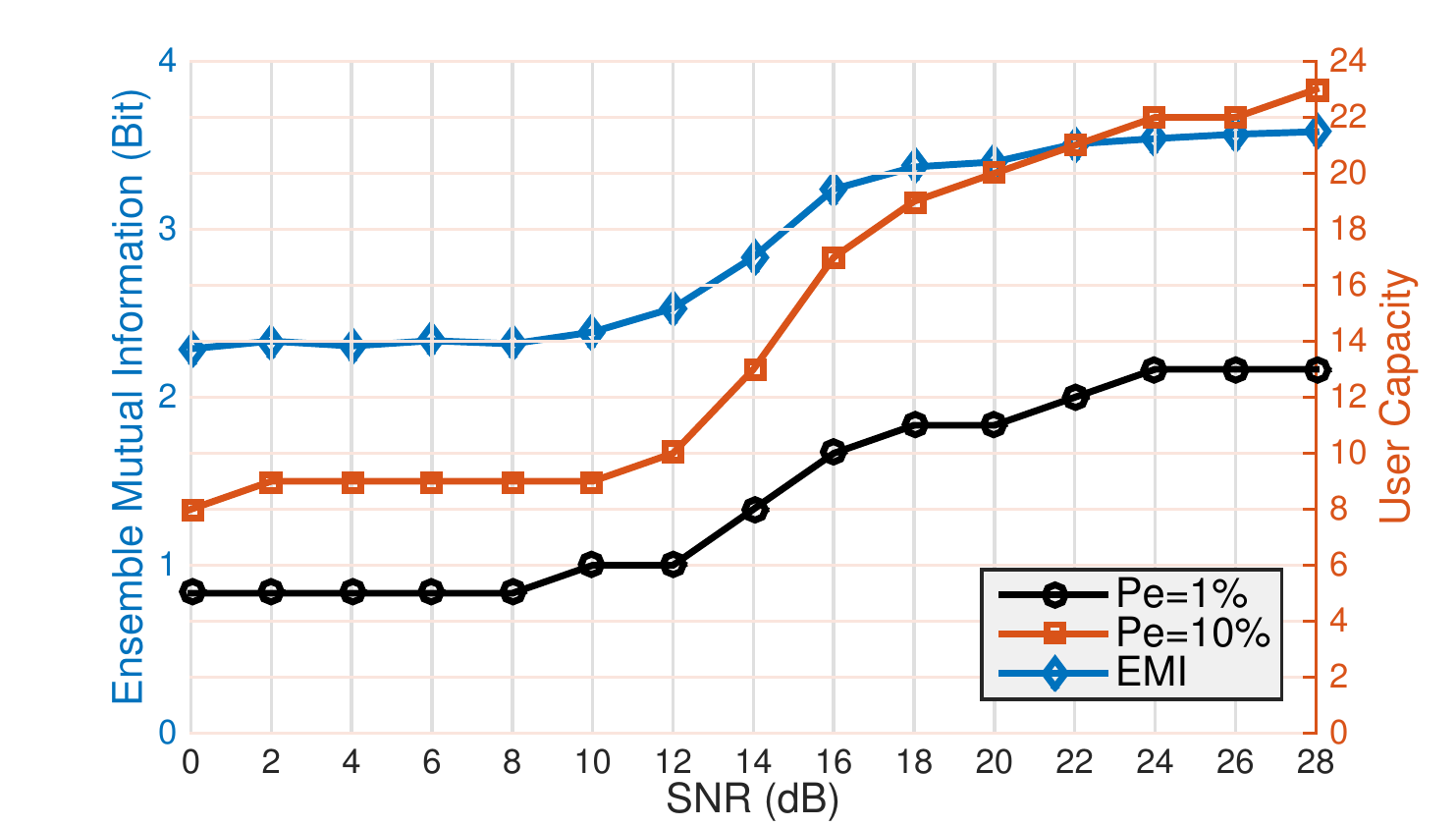}
    \label{fig_SNR}}
  \subfigure[]{
    \includegraphics [width=1.6in] {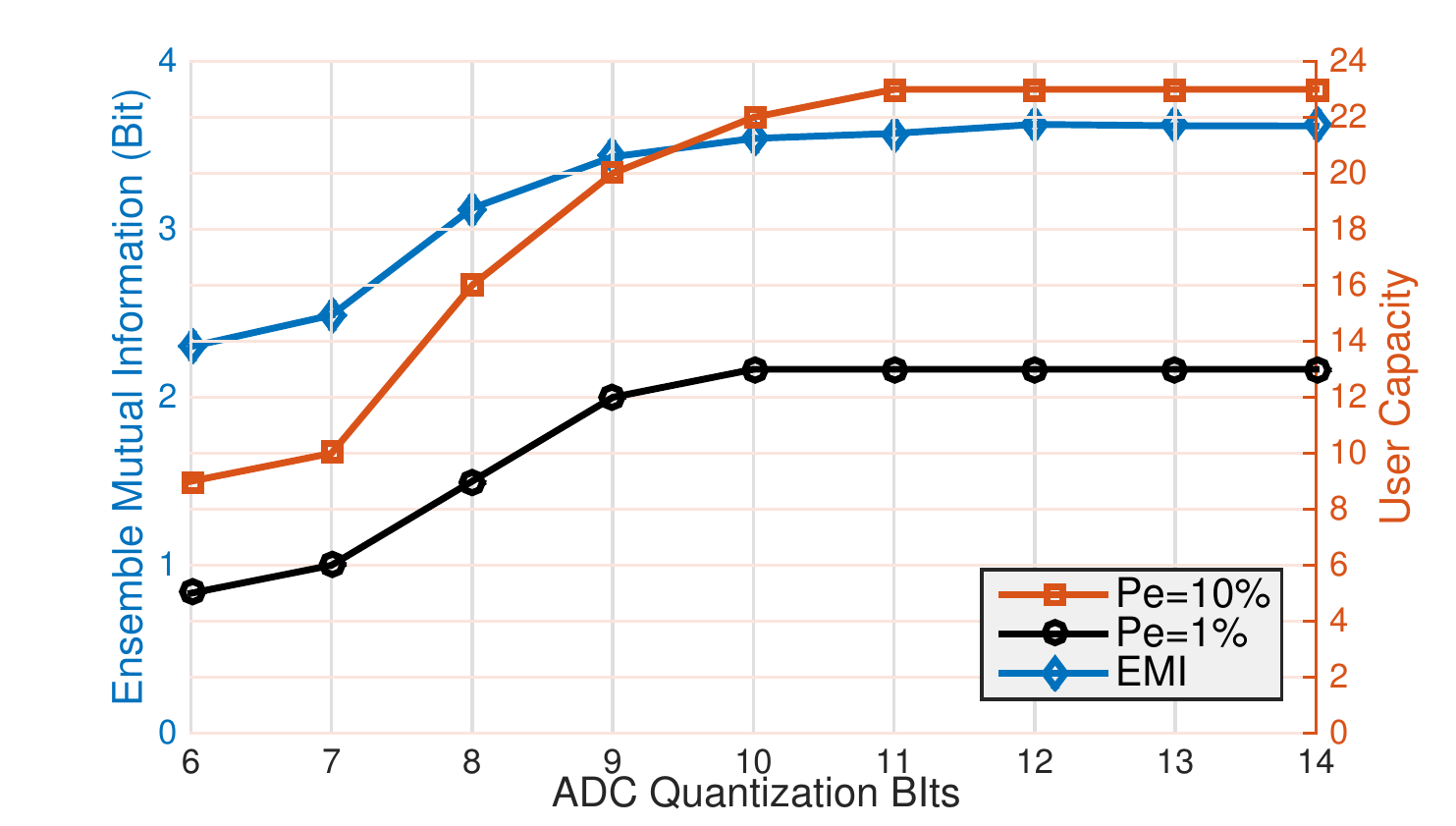}
    \label{fig_ADC}}
  \caption{ (a) User capacity under different RFF noise levels.  (b) User capacity under different ADC quantization bits.}\vspace{-10pt}
     \label{fig_min_d}
\end{figure}

We firstly present user capacity due to various the noise effects in Fig.\ref{fig_min_d}. Here, we set the other parameters as $f_s=4MS/s$, $N_{FFT}=512p$. As the modeling in equation (\ref{eq: ADC}), the noise level of RFF feature is mainly contributed by the ADC quantization error and in-band AWGN noise. The noise level  significantly affects the classification performance of WPLI resulting in the decrease of user capacity we finally obtained. We fix the ADC quantization bit $Q=14$bits and simulate the noise feature value within  SNR$=0\sim 28$dB. The user capacity results are presented in Fig.\ref{fig_SNR}. The EMI and user capacity are decreased synchronously as the AWGN SNR level decreases. It should be noted that, according to equation (\ref{eq: up_UC}), the user capacity we obtained is the upper bound for all classification methods and classifiers using all these RFF samples. Hence in high SNR situations, the typical classification procedure of WPLI can easily achieve the user capacity quite accurately. However, in the extremely low SNR scenarios, it is hard to use a single method or feature to achieve the upper bound of user capacity. Hence more combined features extracted from RFF for multiple classifiers should be utilized to achieve the upper bound of user capacity, as the work in \cite{brik2008wireless,zanetti2010physical}. Moreover, in the low SNR scenarios, the error in signal acquisition procedure of WPLI can also contribute to worsen the classification performance \cite{suski2008using}, which is out of the scope of this paper and can be discussed in future works. 

Then we fixe the SNR$=29$dB and simulate the feature value of ADC quantization error within $Q=6\sim14bits$. The corresponding user capacity results are presented in Fig.\ref{fig_ADC}. The user capacity becomes stable when the ADC quantization bits are increased to 10 bits. In practical applications, the effects of AWGN noise level usually more significant than the effects of ADC quantization error, while the effects of ADC quantization error can be significant when AWGN SNR level is very high. Here we only can simulate the feature value for 14 bits quantization due to the constrains of USRP daughter board, while in practical, 16 or more quantization bits can also be found in higher standard equipment. With the development of device resolutions, the effects can be kept to minimal, 

\subsubsection{Effects of number of FFT points}

\begin{figure}[!t]
  \centering
    \includegraphics [width=2.5in] {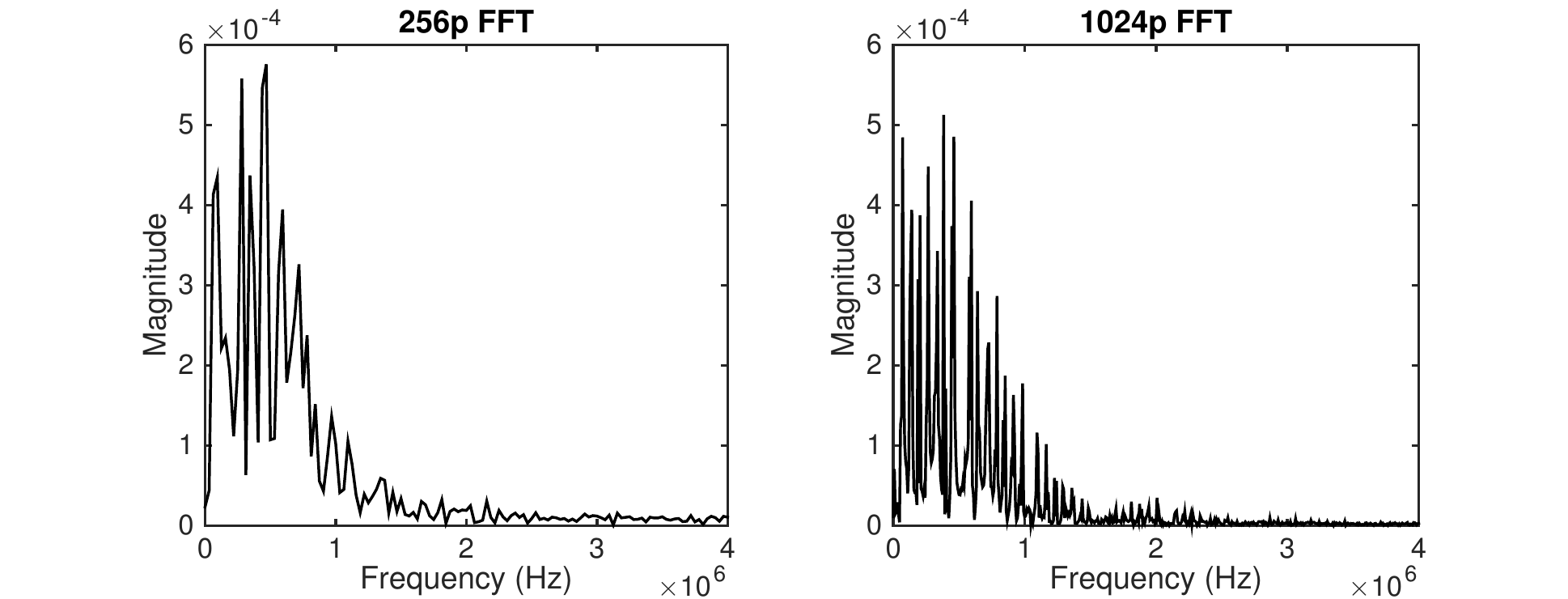}
  \caption{Preamble spectrum of signal under different number of FFT points.}\vspace{-10pt}
     \label{fig_NFFT_PSD}
\end{figure}

\begin{figure}[!t]
  \centering
    \includegraphics [width=2.0in] {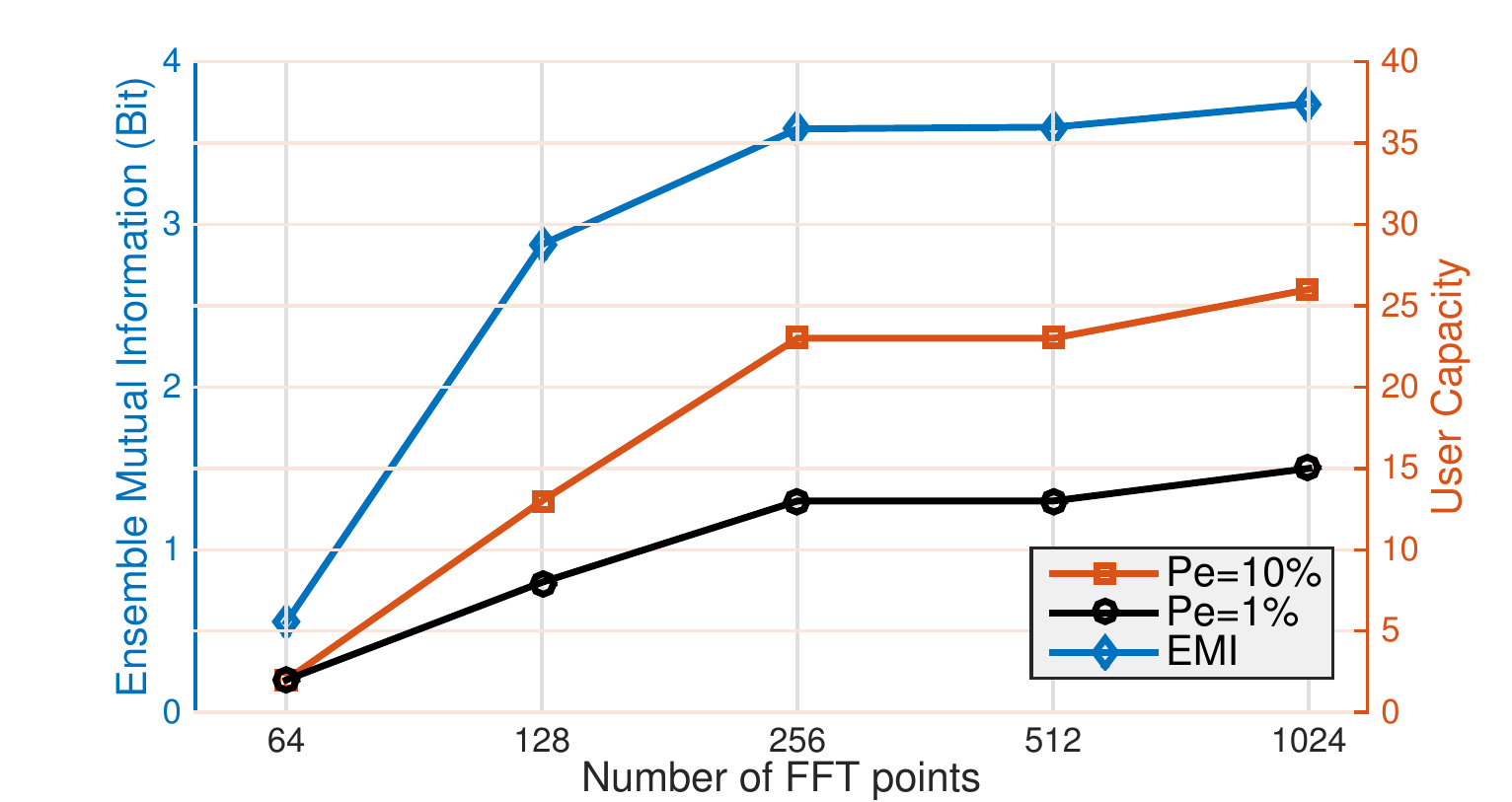}
  \caption{ User capacity under different numbers of FFT points.}\vspace{-15pt}
     \label{fig_NFFT}
\end{figure}

Here we present user capacity due to various number of FFT points within the same sampling rate setting. The number of FFT points $N_{FFT}$ is the key parameter to form the spectral RFF which decides the resolution of the spectral feature $\mathbf{X}$ in equation (\ref{eq: feature}) and consequently reflects the distribution of RFF feature. The specific modeling about the number of FFT points of spectral feature can be found in \cite{danev2009transient}. In Fig.\ref{fig_NFFT_PSD}, we present the preamble spectrum obtained with two different number of FFT points to present difference of resolution. Here, we set the other parameters as $f_s=8$MS/s,  SNR$=29$dB, $Q=14$bits. We simulate the result within number of FFT points $N_{FFT}=64\sim1024$p. The user capacity results are presented in Fig.\ref{fig_NFFT}. From the results, we can see the larger number of FFT points can increase the EMI of feature and improve the performance and user capacity. However, when the resolution is accurate to some extend, the improvement of performance is not so significant. Since the increase of FFT points can cause greater computation burden for WPLI, here involves a trade off for system designer.

\begin{figure}[!t]
  \centering
    \includegraphics [width=2.5in] {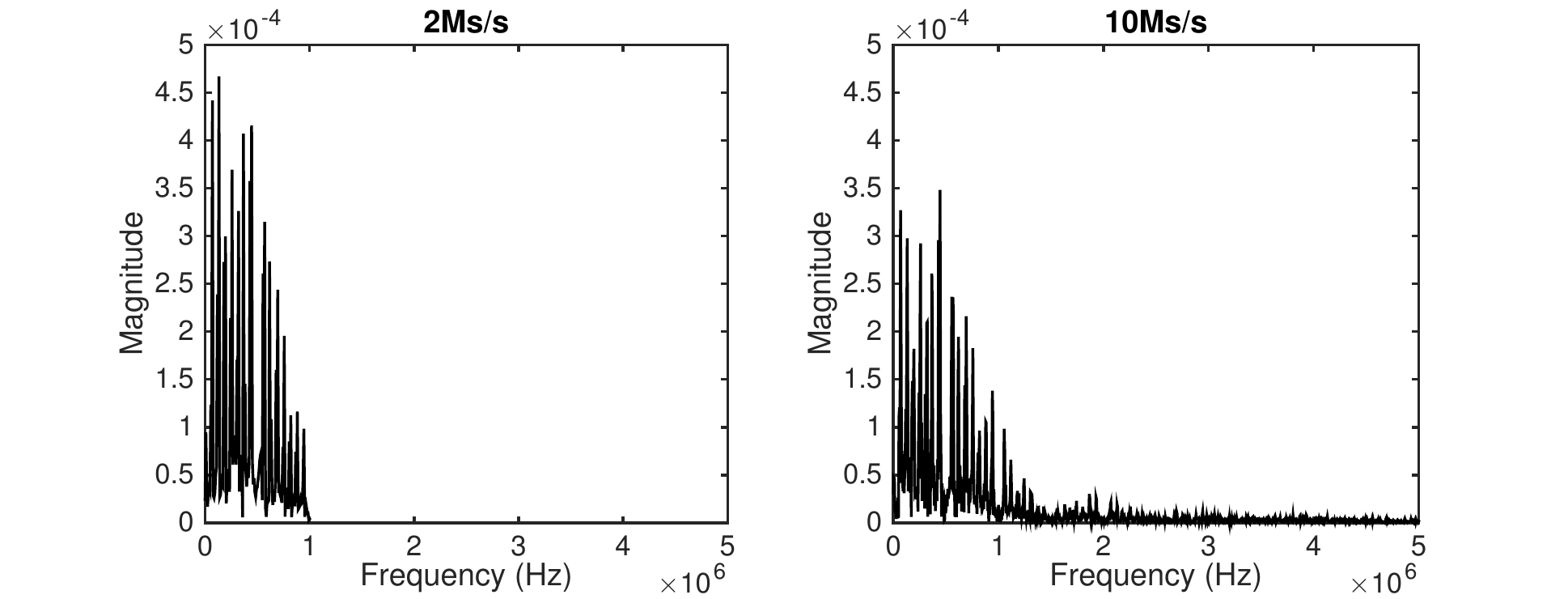}
  \caption{Preamble spectrum of signal under different sampling rates.}\vspace{-10pt}
     \label{fig_Fs_PSD}
\end{figure}

\begin{figure}[!t]
  \centering
    \includegraphics [width=2.0in] {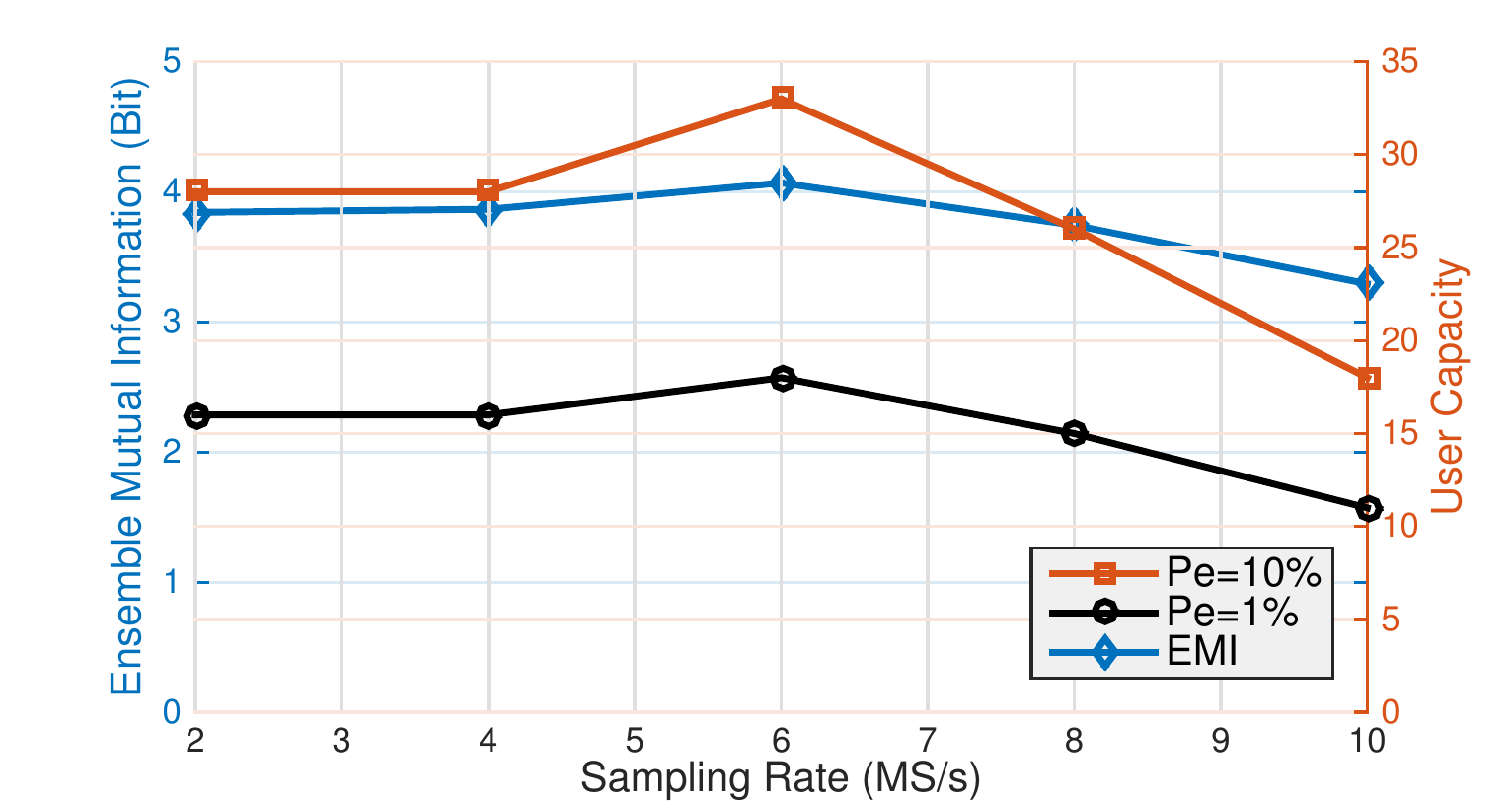}
  \caption{ User capacity under different sampling rates.}\vspace{-15pt}
     \label{fig_fs}
\end{figure}

\subsubsection{Effects of sampling rate}
Here we present user capacity due to various sampling rates of receiver with the same frequency resolution, which are the key parameters to determine the bandwidth of the spectral feature $\mathbf{X}$ in equation (\ref{eq: feature}) and consequently reflects its distribution. In Fig.\ref{fig_Fs_PSD}, we present the preamble spectrum obtained with two different sampling rates to present difference of spectrum bandwidth. In the case $f_s=2$MS/s, the spectrum bandwidth covers the main lobe of signal PSD. While the higher sampling rates can cover more side lobe information of signal PSD which are more beneficial for WPLI performance. However, the choice of sampling rates also involves a trade off that with the increasing of bandwidth, the bandwidth of noise is also increased which can result in the decreasing of signal SNR which worsen the WPLI performance. This phenomenon can be observed in the user capacity characterization and also experimental validations. Here, we set the other parameters as,  SNR$=29$dB for $f_s=8$Ms/s, $Q=14$bits. We simulate the result within different sampling rates, $f_s=2\sim10$MS/s, with the same spectrum resolution ( $N_{FFT}=512p$ for $f_s=4 Ms/s$), the user capacity results of which are presented in Fig.\ref{fig_fs}. It can be inferred that when low-noise devices are applied in high SNR scenarios, a higher sampling rate can be applied for WPLI system. While, for the low-SNR scenarios where low quality devices are applied, a  sampling rate which tightly covers the main lobe of spectrum should be chosen.

\section{Experimental Validations for User Capacity}

In this section, we conduct field experiments on classification error performance according to the different case setting to validate the user capacity characterization in section \ref{sec: app_case}. We still utilize the equipments in section \ref{sec: app_case}. We use newly collected 1000 thousand samples per class to train the LDA training matrix and another 1000 samples per class to be tested for classification. 
For classification procedure of WPLI, we select the Fisher LDA \cite{danev2009transient} as the feature dimensionality reduction technique and the Mahalanobis distance as the distance metric \cite{danev2009physical,zanetti2010physical,danev2009transient}. We set a large LDA subspace dimensionality $\kappa=150$ despite the computation time in order to achieve the optimal classification performance . 

We present the classification error performance of WPLI with selected user number near the upper-bound user capacity we obtained. Hence if the classification error rate is larger than the threshold error rate i.e., $P_e|N_Y > \lambda$, when the user number is larger than the user capacity, i.e., $N_Y > N_C|\lambda$, the user capacity is proved. Meanwhile, we also present the classification error performance when user number is near to the user capacity bound i.e., $N_Y \leq N_C$, to show the tightness of this bound. The classification results are shown in the following figures where the x-axis is the number of test samples, the y-axis is the minimal distance score between test sample and its reference, and the z-axis is the identity number assigned to the test samples. Besides the color of each sample is to present its true identity which can help the reader to compare classified identities of test samples.

\subsection{Effects of RFF noise level}

\begin{figure}[!t]
  \centering
  \subfigure[]{
    \includegraphics [width=1.6in] {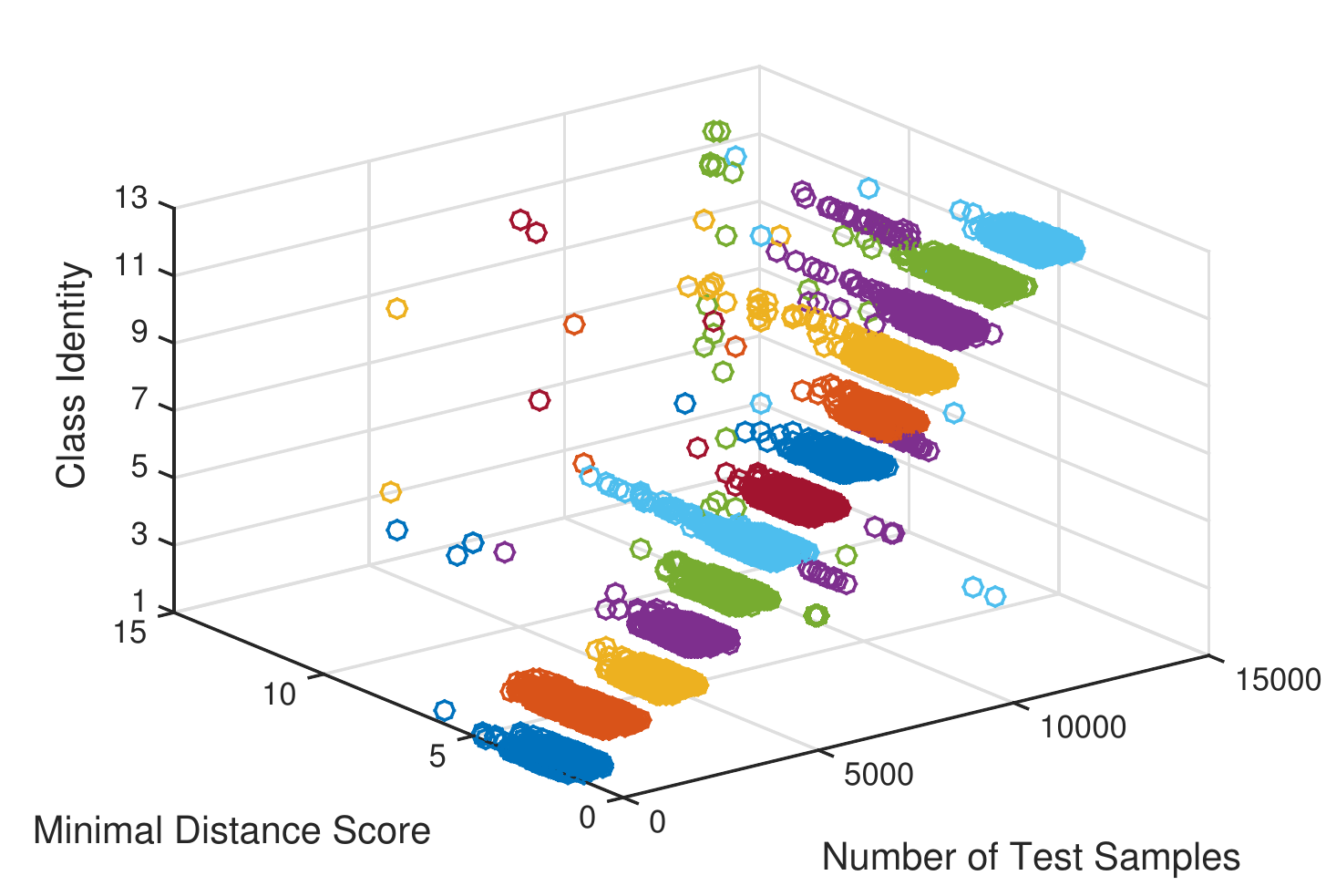}
    \label{fig_class_4M_13c_Pe0_91}}
  \subfigure[]{
    \includegraphics [width=1.6in] {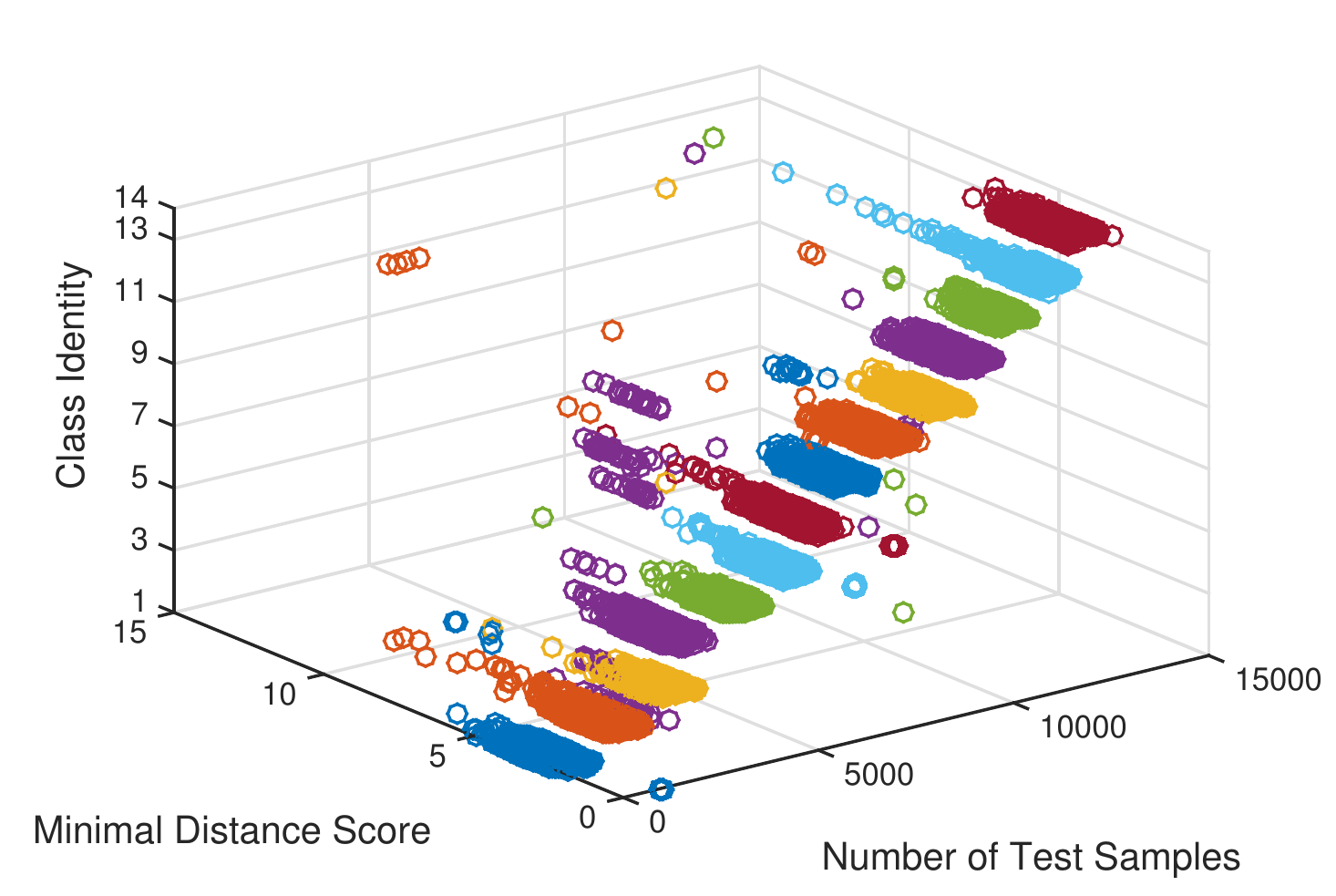}
    \label{fig_class_4M_14c_Pe1_09}}
  \caption{SNR=26dB, $f_s=4MS/s$, $N_{FFT}=512p$. (a) Classification results for 13 classes.  (b)  Classification results for 14 classes.}\vspace{-10pt}
     \label{fig_class_SNR}
\end{figure}

\begin{figure}[!t]
  \centering
  \subfigure[]{
    \includegraphics [width=1.6in] {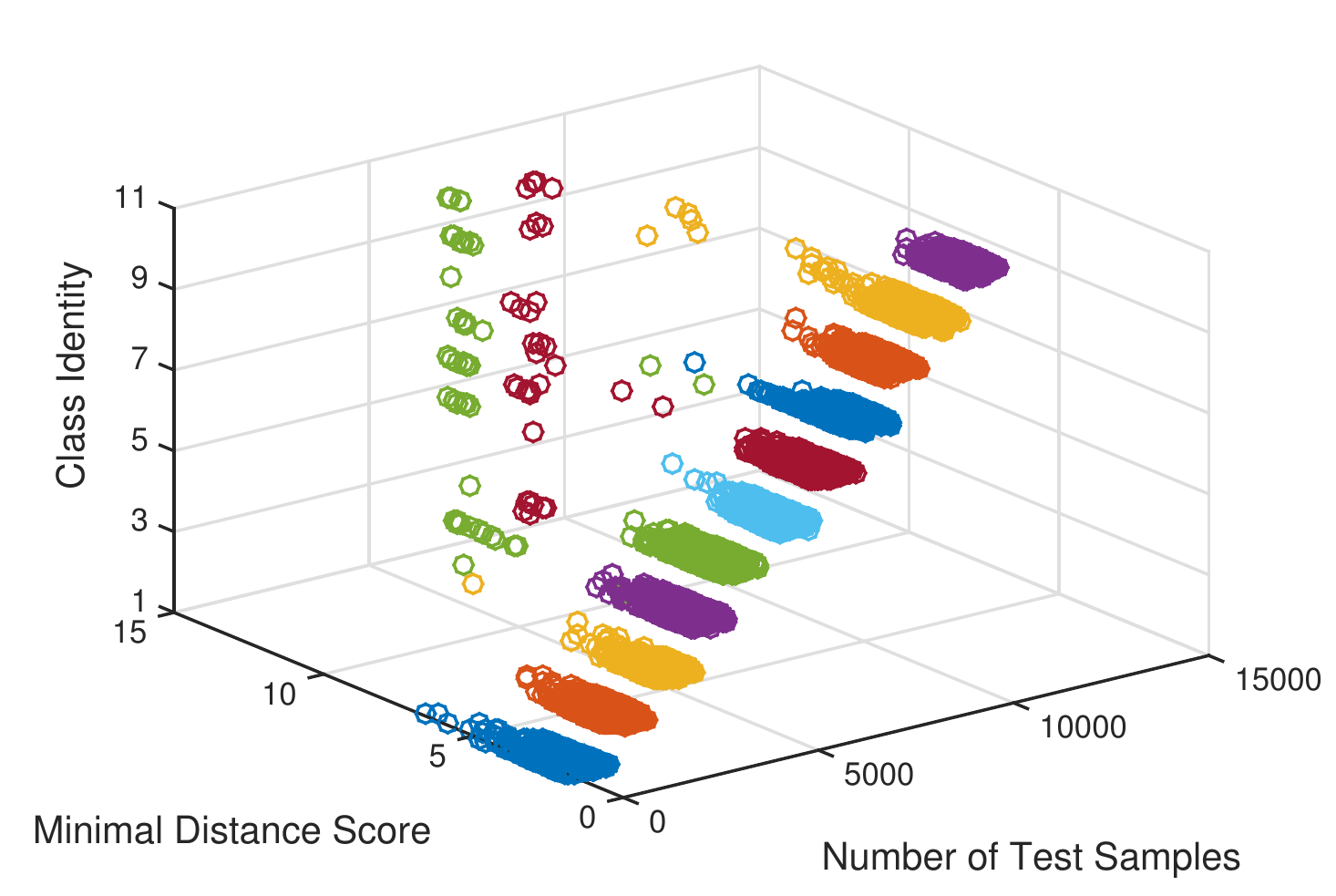}
    \label{fig_class_4M_11c_Pe0_75}}
  \subfigure[]{
    \includegraphics [width=1.6in] {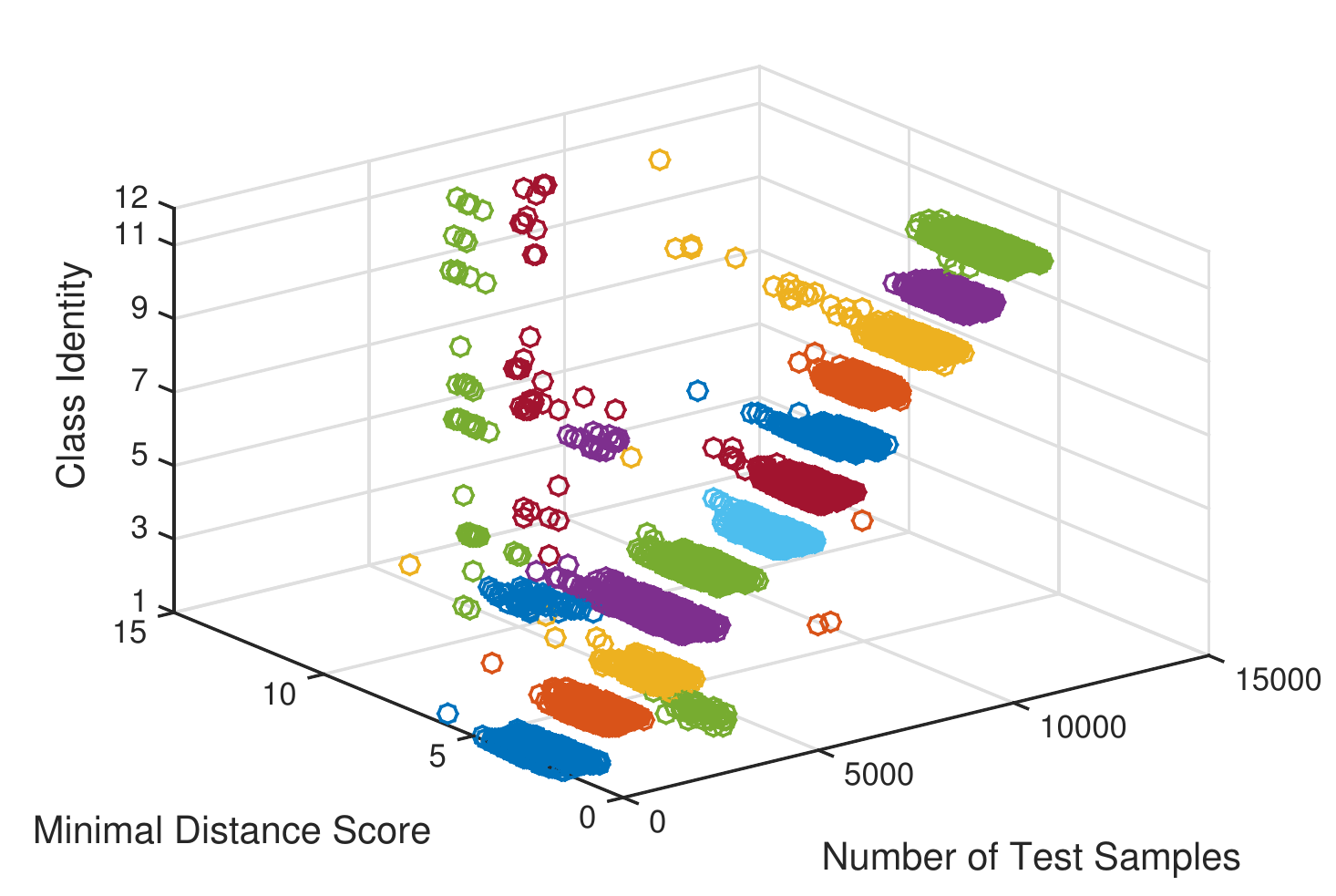}
    \label{fig_class_4M_12c_Pe1_79}}
  \caption{SNR=22dB, $f_s=4MS/s$, $N_{FFT}=512p$. (a) Classification results for 11 classes.  (b) Classification results for 12 classes.}\vspace{-15pt}
     \label{fig_class_SNR_2}
\end{figure}

Here, we use the experiment results to validate the user capacity characterization for RFF noise level case. Because the ADC quantization bits is impossible to change for given hardware setting, we can only fix the ADC quantization bits to Q$=14$bit according to USRP daughter board setting. As the case setting in Fig.\ref{fig_SNR}, we conduct the experiments at SNR=26dB, where the 1\% error rate user capacity is 13, i.e., $N_C=13|P_e\leq1\%$. In Fig.\ref{fig_class_4M_13c_Pe0_91}, classification results of 13 classes, are shown of which the classification error rate is $P_e=0.91\%|N_Y=13$. While in Fig.\ref{fig_class_4M_14c_Pe1_09}, the classification results of 14 classes are shown of which the classification error rate is $P_e=1.09\%|N_Y=14$. Hence we can see the user capacity characterization at this point is validated accurately. Similarly, we change SNR situation and threshold error rate to see validate another point of our user capacity curves. In Fig.\ref{fig_SNR}, when SNR=22dB, the user capacity is 12 with 1\% error rate i.e., $N_C=12|P_e\leq1\%$.  In Fig.\ref{fig_class_4M_11c_Pe0_75}, classification results of 11 classes, are shown of which $P_e=0.75\%|N_Y=11$. While in Fig.\ref{fig_class_4M_12c_Pe1_79}, the classification results of 12 classes are shown of which $P_e=1.79\%|N_Y=12$. The user capacity characterization at this case is slightly larger than the experimental result. As we discuss in the previous section, with the decrease of SNR level, single classifier and single feature selection are not enough achieve the upper bound user capacity.

\subsection{Effects of number of FFT points}

\begin{figure}[!t]
  \centering
  \subfigure[]{
    \includegraphics [width=1.6in] {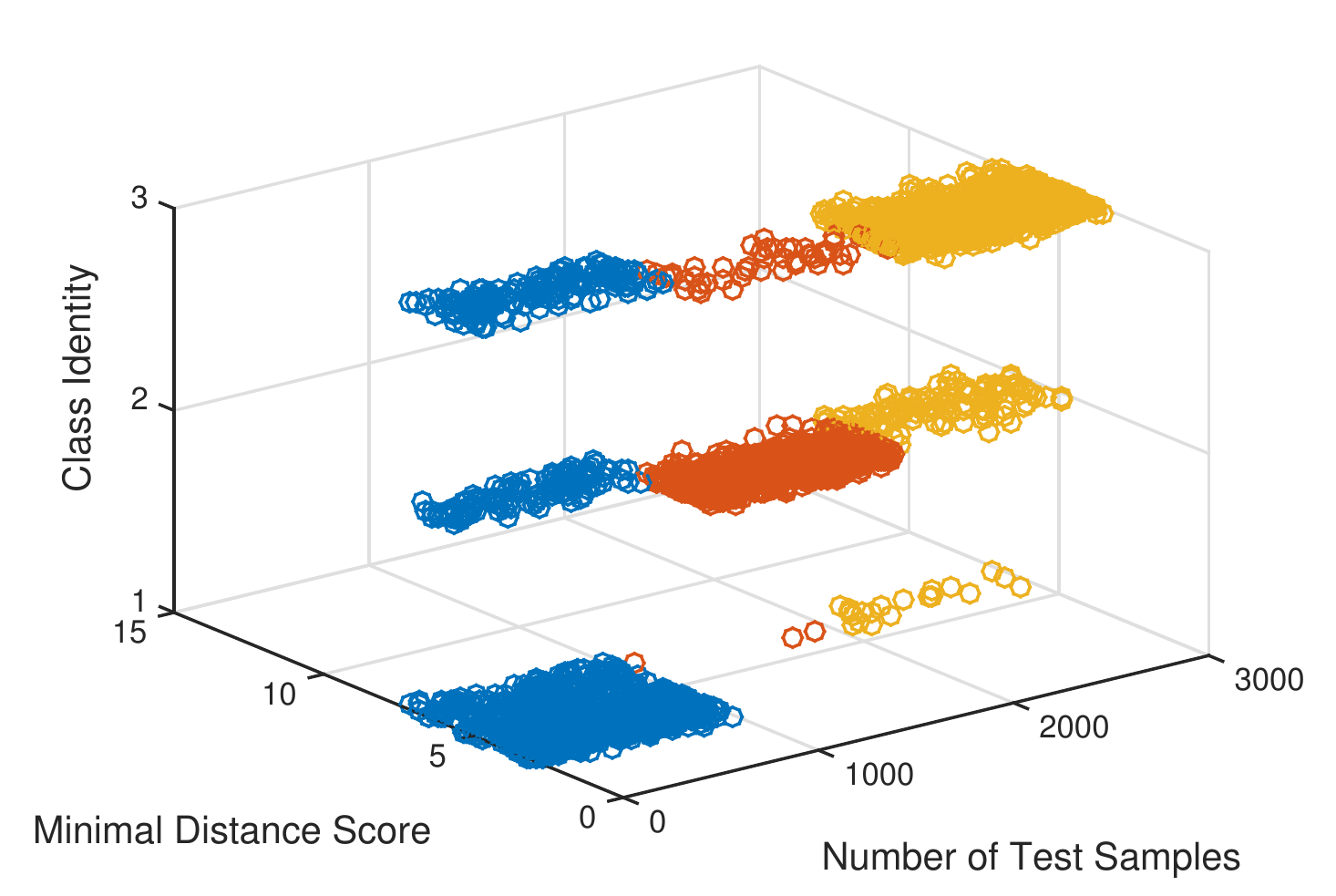}
    \label{fig_class_8M_3c_64}}
  \subfigure[]{
\includegraphics [width=1.6in] {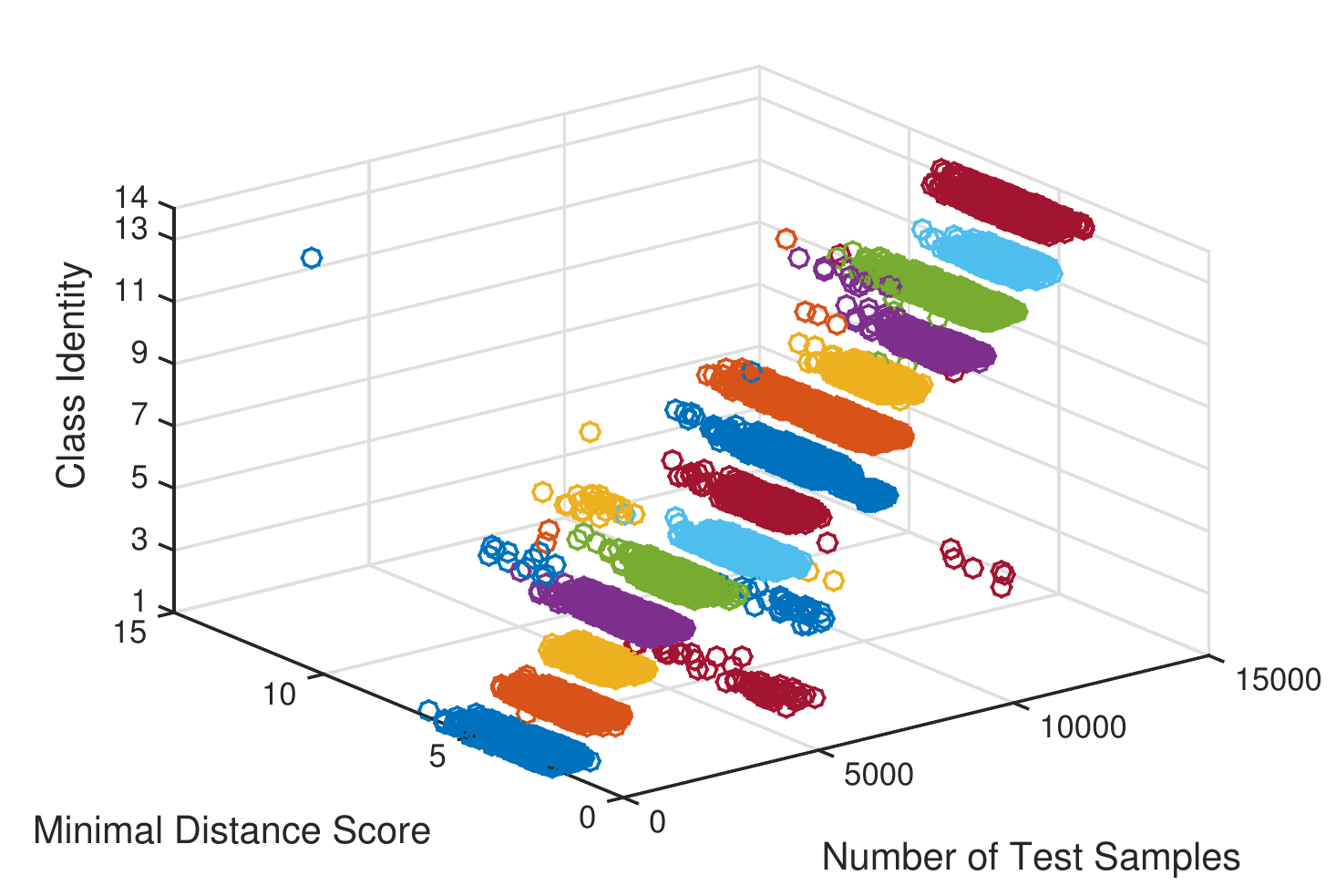}
    \label{fig_class_8M_14c_128}}
      \subfigure[]{
    \includegraphics [width=1.6in] {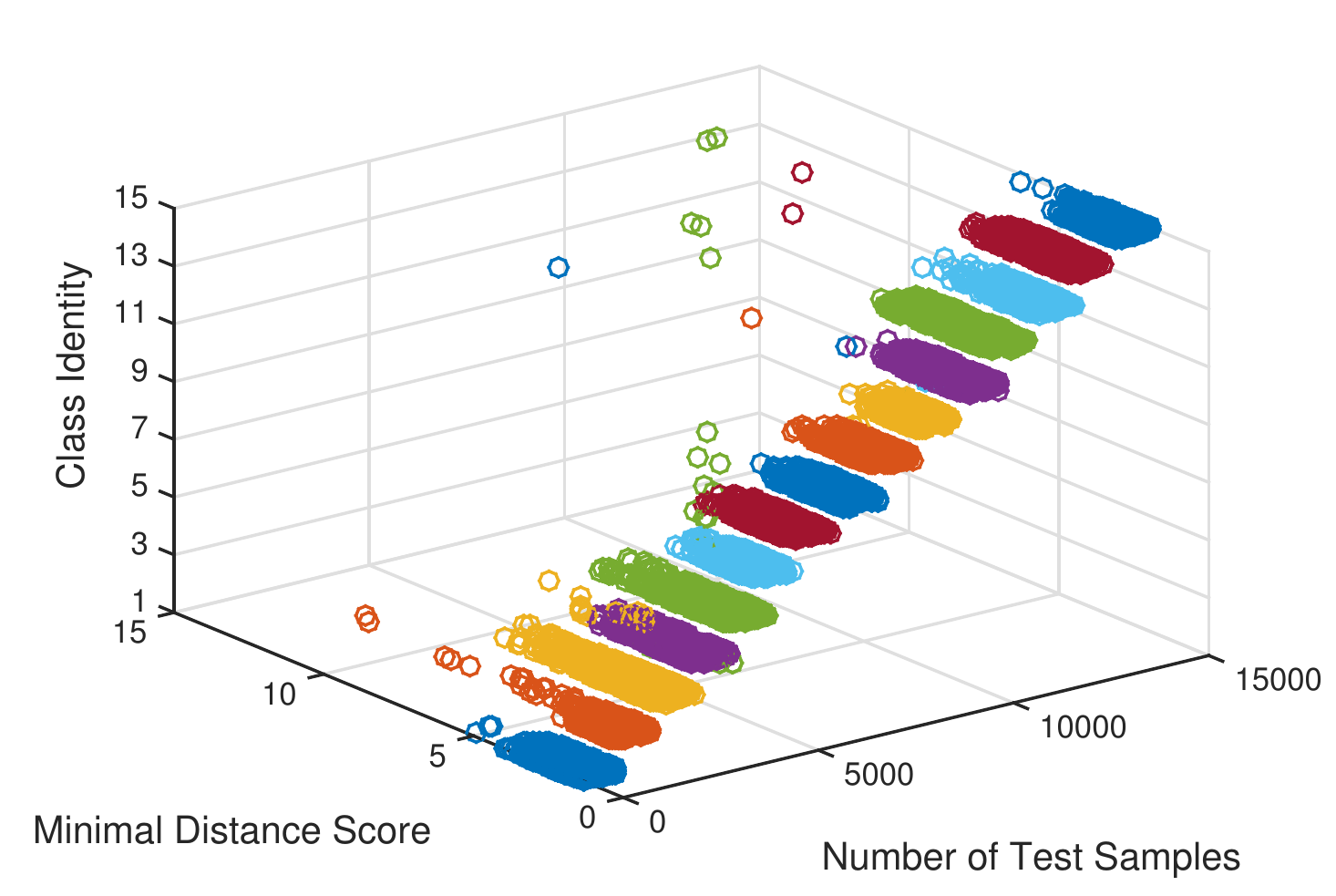}
    \label{fig_class_8M_15c_1024}}
  \caption{SNR=29dB , $f_s=8MS/s$. (a) Classification results for 3 classes, $N_{FFT}=64$.  (b) Classification results  14 classes, $N_{FFT}=128$. (c) Classification results for 15 classes, $N_{FFT}=1024$. }\vspace{-15pt}
     \label{fig_class_NFFT}
\end{figure}

We use the experiment results to validate the user capacity characterization for number of FFT points case. We conduct the experiments as the case setting for Fig.\ref{fig_NFFT}. As the user capacity characterization under $N_{FFT}=64p$ is ,$N_C=2|P_e\leq10\%$. In Fig.\ref{fig_class_8M_3c_64}, the classification results of 3 classes are shown with $P_e=15.67\%|N_Y=3$ which is out of user capacity. For $N_{FFT}=256p$, $N_C=13|P_e\leq1\%$. In Fig.\ref{fig_class_8M_14c_128}, the classification results of 14 classes are shown with $P_e=1.25\%|N_Y=14$ which is still out of user capacity. For $N_{FFT}=1024p$, $N_C=15|P_e\leq1\%$. In Fig.\ref{fig_class_8M_15c_1024}, the classification results of 15 classes are shown with $P_e=0.47\%|N_Y=15$ which is within user capacity. Hence the experimental results match the discussion in previous section very well.

\begin{figure}[!t]
  \centering
  \subfigure[]{
    \includegraphics [width=1.6in] {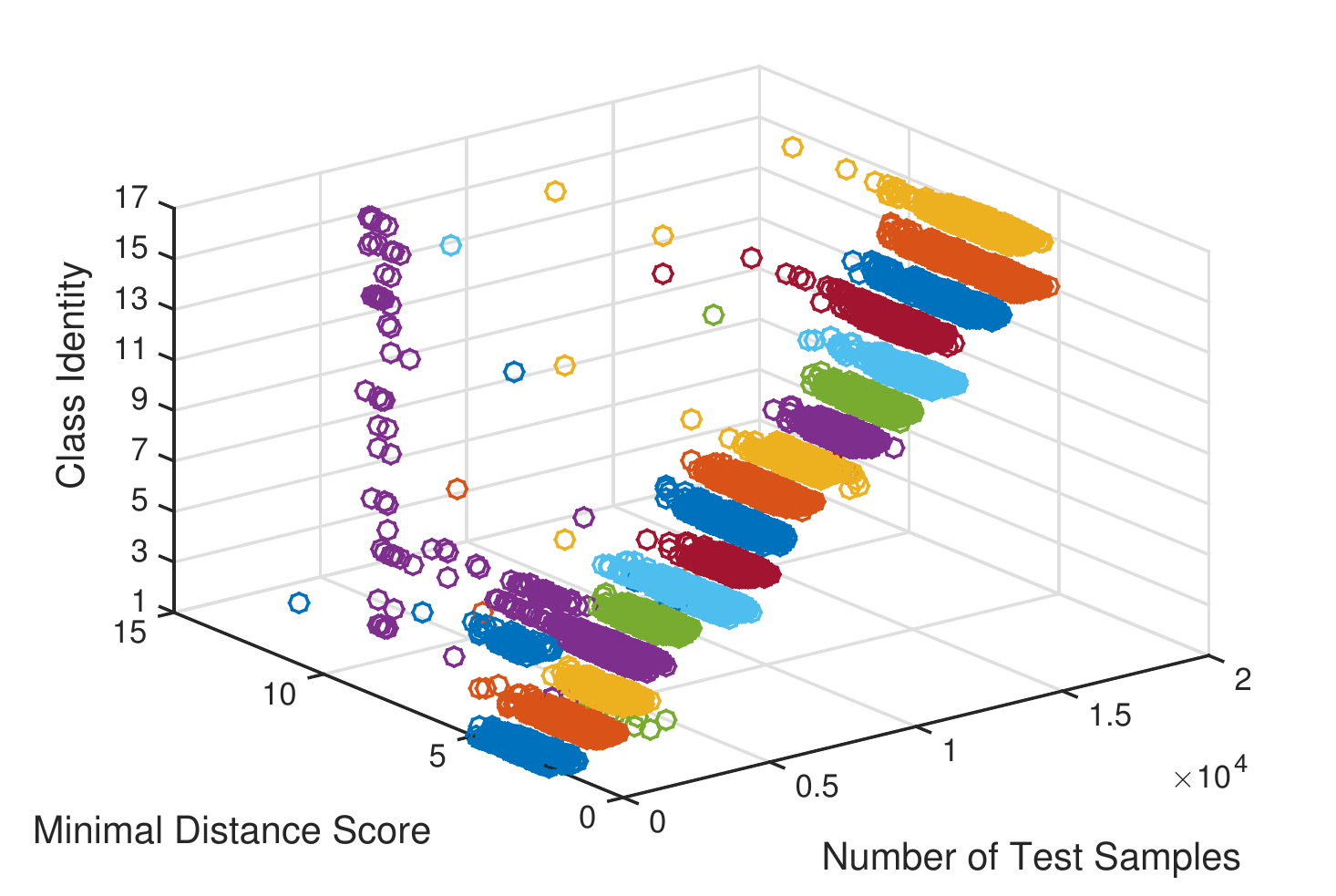}
    \label{fig_class_2M_17c}}
  \subfigure[]{
\includegraphics [width=1.6in] {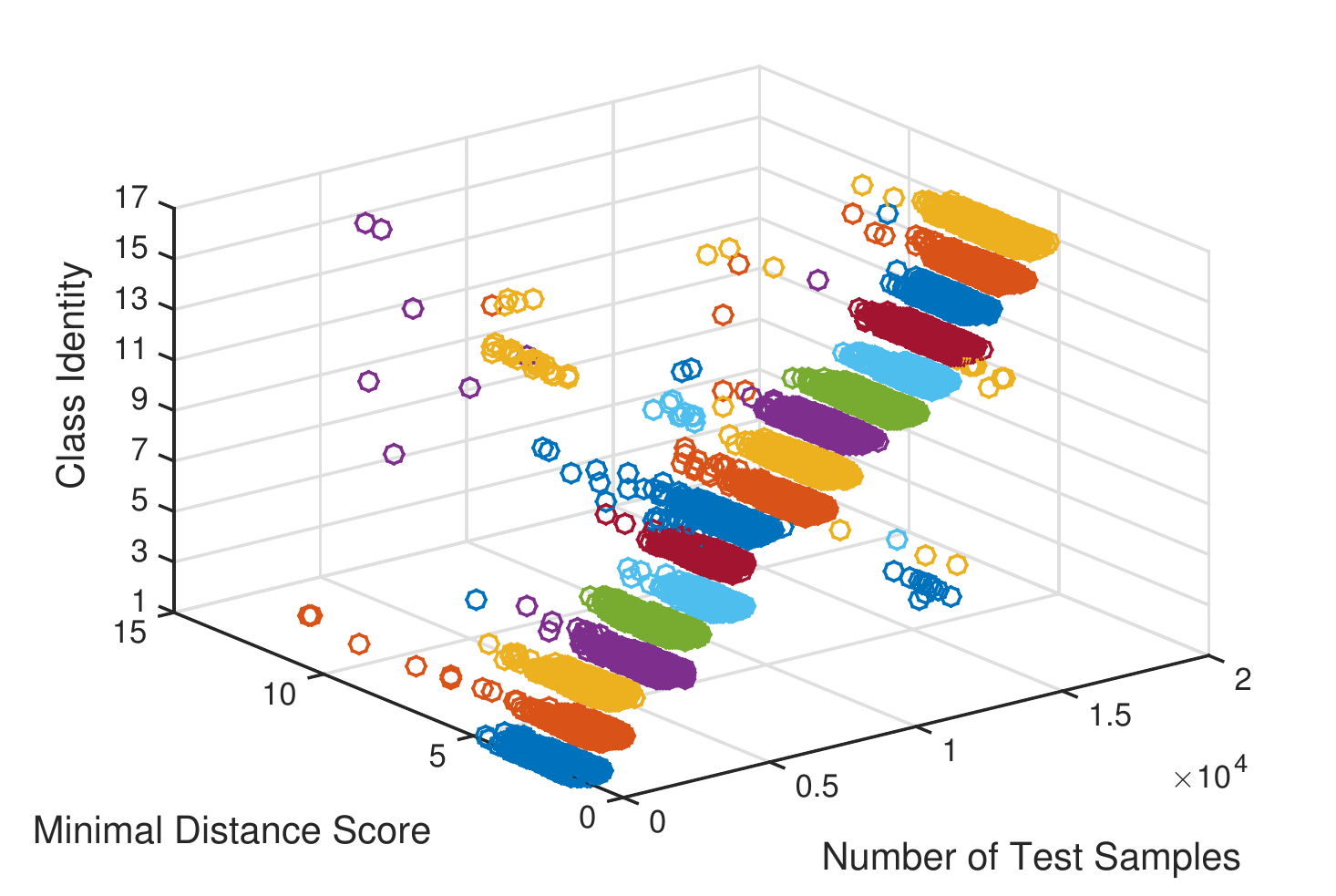}
    \label{fig_class_6M_17c}}
      \subfigure[]{
    \includegraphics [width=1.6in] {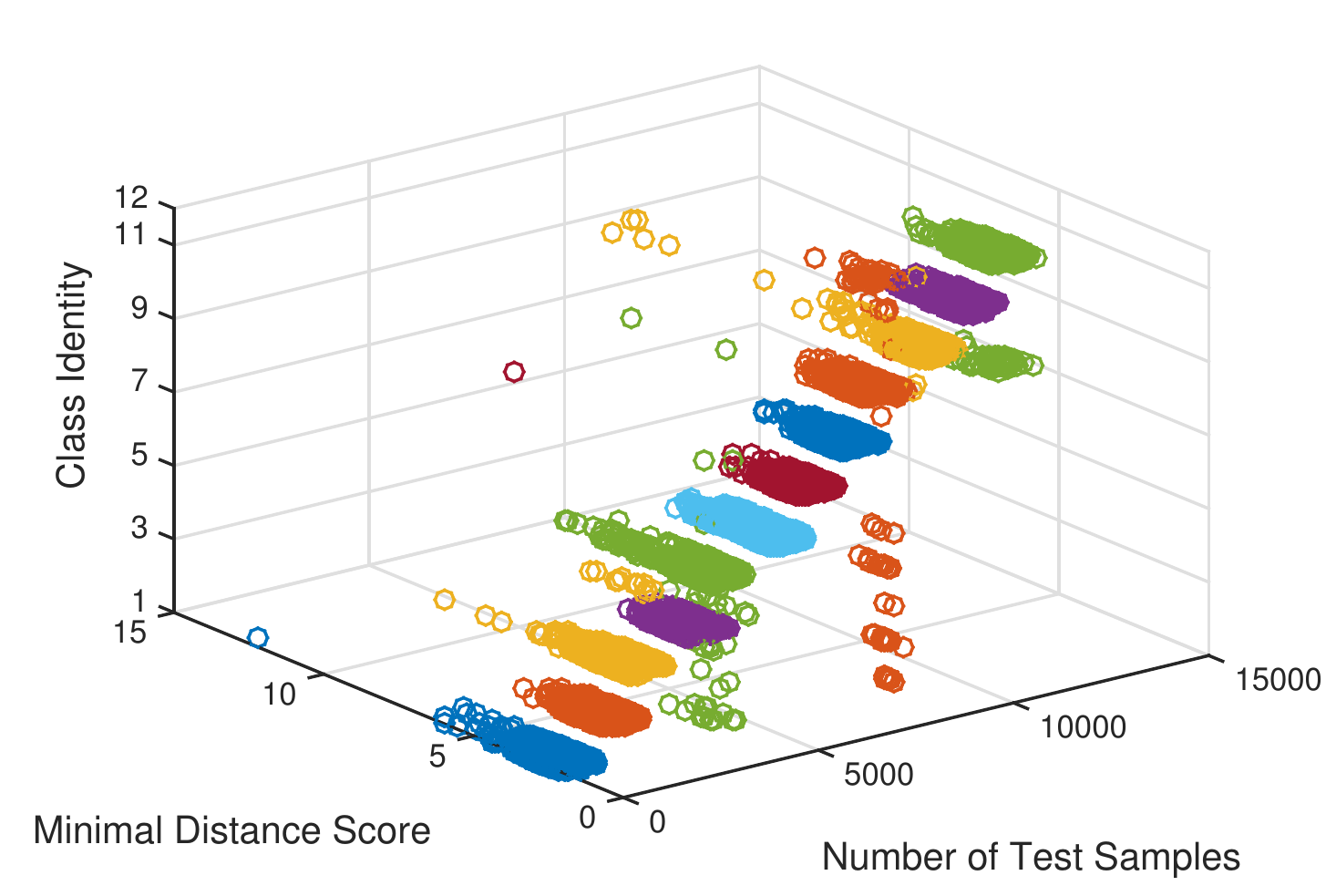}
    \label{fig_class_10M_12c}}
  \caption{SNR=29dB (8MS/s). (a) Classification results for 17 classes, $f_s=2$MS/s.  (b) Classification results  17 classes, $f_s=6$MS/s. (c) Classification results for 12 classes, $f_s=10$MS/s. }\vspace{-15pt}
     \label{fig_class_fs}
\end{figure}

\subsection{Effects of sampling rate}
We use the experiment results to validate the user capacity characterization for sampling rate case. We conduct the experiments as the case setting for Fig.\ref{fig_fs}. As the user capacity characterization under $f_s=2$MS/s is, $N_C=16|P_e\leq1\%$. In Fig.\ref{fig_class_2M_17c}, the classification results of 17 classes are shown with $P_e=1.17\%|N_Y=17$ which is out of user capacity. For $f_s=6$MS/s, $N_C=17|P_e\leq1\%$. In Fig.\ref{fig_class_6M_17c}, the classification results of 17 classes are shown with $P_e=0.58\%|N_Y=17$ which is within the capacity. For $f_s=10$MS/s, $N_C=11|P_e\leq1\%$. In Fig.\ref{fig_class_10M_12c}, the classification results of 12 classes under $f_s=10$MS/s are shown with $P_e=1.51\%|N_Y=12$, which is out of the capacity.  The analyses in our previous discussion is also validated.
\vspace{-5pt}

\section{Conclusion}
\label{sec: C}
In this work, we establish a theoretical understanding on user capacity of Wireless Physical-layer Identification (WPLI) in an information-theoretic perspective unprecedentedly. Specifically, Radio Frequency Fingerprint (RFF) features of WPLI are analyzed in an information-theoretic perspective, including feature selection, extraction, noise level, resolution, and bandwidth, which advance the understanding of RFF feature. We then propose an information-theoretic approach based on mutual information between RFF and user identity to characterize the user capacity of WPLI. Using this theoretical tool, the achievable user capacity under practical constrains of a WPLI system is characterized with data collected by off-the-shelf receiving devices. Various experiments on classification error performance of a practical system are conducted to validate the accuracy and tightness of the information-theoretic user capacity characterization.
\vspace{-5pt}

\bibliographystyle{IEEEtran}
\bibliography{User_capacity.bib}

\begin{thebibliography}{10}
\providecommand{\url}[1]{#1}
\csname url@samestyle\endcsname
\providecommand{\newblock}{\relax}
\providecommand{\bibinfo}[2]{#2}
\providecommand{\BIBentrySTDinterwordspacing}{\spaceskip=0pt\relax}
\providecommand{\BIBentryALTinterwordstretchfactor}{4}
\providecommand{\BIBentryALTinterwordspacing}{\spaceskip=\fontdimen2\font plus
\BIBentryALTinterwordstretchfactor\fontdimen3\font minus
  \fontdimen4\font\relax}
\providecommand{\BIBforeignlanguage}[2]{{%
\expandafter\ifx\csname l@#1\endcsname\relax
\typeout{** WARNING: IEEEtran.bst: No hyphenation pattern has been}%
\typeout{** loaded for the language `#1'. Using the pattern for}%
\typeout{** the default language instead.}%
\else
\language=\csname l@#1\endcsname
\fi
#2}}
\providecommand{\BIBdecl}{\relax}
\BIBdecl

\bibitem{danev2012physical}
B.~Danev, D.~Zanetti, and S.~Capkun, ``On physical-layer identification of
  wireless devices,'' \emph{ACM Computing Surveys (CSUR)}, vol.~45, no.~1,
  p.~6, 2012.

\bibitem{danev2009physical}
B.~Danev, T.~S. Heydt-Benjamin, and S.~Capkun, ``Physical-layer identification
  of rfid devices.'' in \emph{Usenix Security Symposium}, 2009, pp. 199--214.

\bibitem{polak2011identifying}
A.~C. Polak, S.~Dolatshahi, and D.~L. Goeckel, ``Identifying wireless users via
  transmitter imperfections,'' \emph{Selected Areas in Communications, IEEE
  Journal on}, vol.~29, no.~7, pp. 1469--1479, 2011.

\bibitem{polak2011rf}
A.~C. Polak and D.~L. Goeckel, ``Rf fingerprinting of users who actively mask
  their identities with artificial distortion,'' in \emph{Signals, Systems and
  Computers (ASILOMAR), 2011 Conference Record of the Forty Fifth Asilomar
  Conference on}.\hskip 1em plus 0.5em minus 0.4em\relax IEEE, 2011, pp.
  270--274.

\bibitem{liu2008specific}
M.-W. Liu and J.~F. Doherty, ``Specific emitter identification using nonlinear
  device estimation,'' in \emph{Sarnoff Symposium, 2008 IEEE}.\hskip 1em plus
  0.5em minus 0.4em\relax IEEE, 2008, pp. 1--5.

\bibitem{zanetti2010physical}
D.~Zanetti, B.~Danev \emph{et~al.}, ``Physical-layer identification of uhf rfid
  tags,'' in \emph{Proceedings of the sixteenth annual international conference
  on Mobile computing and networking}.\hskip 1em plus 0.5em minus 0.4em\relax
  ACM, 2010, pp. 353--364.

\bibitem{jana2010fast}
S.~Jana and S.~K. Kasera, ``On fast and accurate detection of unauthorized
  wireless access points using clock skews,'' \emph{Mobile Computing, IEEE
  Transactions on}, vol.~9, no.~3, pp. 449--462, 2010.

\bibitem{brik2008wireless}
V.~Brik, S.~Banerjee, M.~Gruteser, and S.~Oh, ``Wireless device identification
  with radiometric signatures,'' in \emph{Proceedings of the 14th ACM
  international conference on Mobile computing and networking}.\hskip 1em plus
  0.5em minus 0.4em\relax ACM, 2008, pp. 116--127.

\bibitem{danev2009transient}
B.~Danev and S.~Capkun, ``Transient-based identification of wireless sensor
  nodes,'' in \emph{Proceedings of the 2009 International Conference on
  Information Processing in Sensor Networks}.\hskip 1em plus 0.5em minus
  0.4em\relax IEEE Computer Society, 2009, pp. 25--36.

\bibitem{scanlon2010feature}
P.~Scanlon, I.~O. Kennedy, and Y.~Liu, ``Feature extraction approaches to rf
  fingerprinting for device identification in femtocells,'' \emph{Bell Labs
  Technical Journal}, vol.~15, no.~3, pp. 141--151, 2010.

\bibitem{brown2009information}
G.~Brown, ``An information theoretic perspective on multiple classifier
  systems,'' in \emph{Multiple Classifier Systems}.\hskip 1em plus 0.5em minus
  0.4em\relax Springer, 2009, pp. 344--353.

\bibitem{cover2012elements}
T.~M. Cover and J.~A. Thomas, \emph{Elements of information theory}.\hskip 1em
  plus 0.5em minus 0.4em\relax John Wiley \& Sons, 2012.

\bibitem{suski2008using}
W.~C. Suski, M.~A. Temple, M.~J. Mendenhall, and R.~F. Mills, ``Using spectral
  fingerprints to improve wireless network security,'' in \emph{Global
  Telecommunications Conference, 2008. IEEE GLOBECOM 2008. IEEE}.\hskip 1em
  plus 0.5em minus 0.4em\relax IEEE, 2008, pp. 1--5.

\bibitem{barbeau2006detection}
M.~Barbeau, J.~Hall, and E.~Kranakis, ``Detection of rogue devices in bluetooth
  networks using radio frequency fingerprinting,'' in \emph{Proceedings of the
  3rd IASTED International Conference on Communications and Computer Networks,
  CCN}.\hskip 1em plus 0.5em minus 0.4em\relax Citeseer, 2006, pp. 4--6.

\bibitem{zhou2010multi}
Z.-H. Zhou and N.~Li, ``Multi-information ensemble diversity,'' in
  \emph{Multiple Classifier Systems}.\hskip 1em plus 0.5em minus 0.4em\relax
  Springer, 2010, pp. 134--144.

\bibitem{ozertem2005detection}
U.~Ozertem, D.~Erdogmus, and I.~Santamaria, ``Detection of nonlinearly
  distorted signals using mutual information,'' in \emph{13th European Signal
  Processing Conference (EUSIPCO 2005)}.\hskip 1em plus 0.5em minus 0.4em\relax
  Citeseer, 2005.

\bibitem{ozertem2006spectral}
U.~Ozertem, D.~Erdogmus, and R.~Jenssen, ``Spectral feature projections that
  maximize shannon mutual information with class labels,'' \emph{Pattern
  Recognition}, vol.~39, no.~7, pp. 1241--1252, 2006.

\bibitem{fasshauerkernel}
G.~Fasshauer and M.~McCourt, ``Kernel-based approximation methods using
  matlab,'' 2015.

\end{thebibliography}

\end{document}